%% file: McMahonDeGreve_arXiv.tex
\documentclass[11pt]{article}
\usepackage[skip=10pt,font=small]{caption} % set figure captions to use font size \small (which is apparently what the main text is set in)
\usepackage{floatrow}
\DeclareFloatFont{tablefontsize}{\footnotesize}% define font size for tables
\usepackage{ragged2e} % provides \raggedright, which is needed to turn off the default \centering that floatrow causes

\usepackage[letterpaper,left=1.55in,right=1.55in,top=1in,bottom=1.25in]{geometry} % setup page margins

\usepackage{xkeyval}
\usepackage{xifthen}
\usepackage{titlesec,titletoc}
\usepackage{textcase} % provides \MakeTextUppercase and \MakeTextLowercase

\usepackage[osf,sc]{mathpazo} % use Palatino for mathematics
\usepackage[scaled=0.90]{helvet} % Helvetica
\usepackage[scaled=0.85]{beramono} % Bera Monospaced
\usepackage[T1]{fontenc}
\usepackage{textcomp}

\usepackage[usenames,dvipsnames,svgnames]{xcolor}

\usepackage[hyperref=true,
            url=false,
            isbn=false,
						pageranges=false,
            backref=true,
						style=phys-plm,
            maxcitenames=3,
            maxbibnames=100,
            block=none,
						sorting=none,
						backend=bibtex]{biblatex}
\renewcommand*{\bibfont}{\small} % use \small font for bibliography

\usepackage[bottom]{footmisc}% places footnotes at page bottom
\usepackage{mathrsfs,amsfonts,amssymb,amsmath,graphicx,multirow,xcolor,textcomp,gensymb}
\usepackage[utf8]{inputenc} % ensures that special characters (for example, in the bibliography) will work
\usepackage[log-declarations=false]{xparse}
\usepackage{flafter} % make sure figures do not appear before their text
\usepackage{siunitx} % SI units
\usepackage{booktabs} % better tables
\usepackage{commath} % \abs{}
\allowdisplaybreaks % use this so that in a \begin{align} region, equations can be spread over multiple pages
\usepackage{authblk} % for author affiliations

\include{CustomCommands} % PLM custom math and other commands

\begin{document}

\title{Towards Quantum Repeaters with Solid-State Qubits: Spin-Photon Entanglement Generation using Self-Assembled Quantum Dots}

\author[1]{\normalsize Peter L. McMahon\thanks{pmcmahon@stanford.edu}}
\author[2]{\normalsize Kristiaan De Greve\thanks{kdegreve@physics.harvard.edu}}

\affil[1]{\footnotesize E.\,L. Ginzton Laboratory and Department of Applied Physics, Stanford University, Stanford, California 94305, USA}
\affil[2]{\footnotesize Department of Physics, Harvard University, Cambridge, Massachusetts 02138, USA}

\date{\vspace{4ex}{\footnotesize This is an expanded version of a chapter to appear in:\\{\it Engineering the Atom-Photon Interaction} (Springer-Verlag, 2015)}\vspace{1ex}}

\maketitle

\abstract{In this chapter we review the use of spins in optically-active InAs quantum dots as the key physical building block for constructing a quantum repeater, with a particular focus on recent results demonstrating entanglement between a quantum memory (electron spin qubit) and a flying qubit (polarization- or frequency-encoded photonic qubit). This is a first step towards demonstrating entanglement between distant quantum memories (realized with quantum dots), which in turn is a milestone in the roadmap for building a functional quantum repeater. We also place this experimental work in context by providing an overview of quantum repeaters, their potential uses, and the challenges in implementing them.}

%------------------------------------------------------------------------------------

% do not indent first line of each paragraph; instead put in a full line break
% MUST come after {abstract}, which resets the parskip and parindent parameters
\setlength{\parskip}{10pt plus 1pt minus 1pt}
\setlength{\parindent}{0pt}

\section{Introduction}

Self-assembled InAs quantum dots\footnote{This chapter focuses exclusively on optically-active self-assembled quantum dots, which can trap single charges (electrons or holes), as well as neutral and charged excitons, due to the difference in the bandgap of the QD material versus that of the surrounding host material. References \cite{kiravittaya_advanced_2009,buckley_engineered_2012,warburton_single_2013,de_greve_ultrafast_2013} provide detailed reviews of how these quantum dots are formed, how they provide a photonic interface, and how they can store spin qubits. This chapter does not review any of the work in the electrostatically-defined quantum dot \cite{hanson_spins_2007} community, which generally combines bandgap discontinuities in one dimension with potentials formed by the application of a voltage over gate electrodes  to trap charges in the other two dimensions (in these devices, either electrons or holes are trapped, but not both at the same time). We use the shorthand ``quantum dots'' in this chapter to refer exclusively to optically-active, self-assembled quantum dots.} can trap a single electron; when the quantum dot is in an external magnetic field, a trapped electron's spin states can be used to encode a quantum bit (qubit). Over the past decade, a series of studies \cite{de_greve_ultrafast_2013} have shown that such a qubit can be optically initialized \cite{atature_quantum-dot_2006,xu_fast_2007}, controlled \cite{press_complete_2008,berezovsky_picosecond_2008} and measured \cite{press_complete_2008,berezovsky_nondestructive_2006,atature_observation_2007}. Measurements of the coherence time of such a qubit have shown that the time required to perform an arbitrary single qubit operation ($\sim 50 \ps$ \cite{press_complete_2008}) on the qubit is roughly five orders of magnitude shorter than the spin echo $T_2$ time ($\sim 3 \us$ \cite{greilich_mode_2006,press_ultrafast_2010}). In light of this, electron spins in quantum dots\footnote{Incidentally, holes can also be trapped in quantum dots, and the (pseudo-)spin of the hole can also be used as a qubit. Analogous demonstrations to those performed with electron spins have been done with hole spins, including: initialization \cite{gerardot_optical_2008,ramsay_fast_2008,de_greve_ultrafast_2011}, complete control \cite{de_greve_ultrafast_2011,greilich_optical_2011}, optical readout \cite{de_greve_ultrafast_2011,greilich_optical_2011}, $T_2^*$ measurement \cite{brunner_coherent_2009,de_greve_ultrafast_2011,greilich_optical_2011}, and $T_2$ (spin echo) measurement \cite{de_greve_ultrafast_2011}. Hole spin qubits in InAs QDs have an advantage over electron spin qubits in InAs QDs: they have a much-reduced hyperfine interaction with the nuclear spin ensemble in the QD, and this results in hole spin qubits exhibiting non-hysteretic behaviour, whereas electron spin qubits suffer from a pronounced hysteresis \cite{de_greve_ultrafast_2011}. In other words, control of electron spin qubits in this material system depends on the history of previous operations performed on it, whereas control of hole spin qubits does not require knowledge of the history; this is a significant difference when long sequences of operations are to be used.} are considered appealing candidates as quantum memories, and will be even more so if dynamical decoupling techniques \cite{viola_dynamical_1999,khodjasteh_fault-tolerant_2005} can be used to further extend the coherence time\footnote{The coherence time of the quantum memory plays an important role in both experiments demonstrating entanglement distribution, and in the design and implementation of quantum repeaters. We briefly discuss the constraints imposed by the coherence time in Section \ref{sec:T2timeConstraints}.}. Long-distance quantum cryptography will likely require the development of quantum repeaters, as will other applications of remote entangled states. Charged quantum dots are an interesting candidate technology for building quantum repeaters, because they provide both a stationary qubit (to be used as a memory), and a fast optical interface.\footnote{The requirements for building a useful quantum network with quantum repeaters are exceptionally challenging, and no candidate technologies at present offer a clear path towards implementation of practical quantum repeaters. Quantum dots do however offer many of the basic features that are required to implement a repeater, and are good candidates for developing small-scale demonstrations of some of the key parts of a quantum network.} One of the very first steps towards building a quantum repeater using quantum dots is to show that one can generate a photonic qubit that is entangled with a spin (memory) qubit.

In this chapter, we review how quantum dots may be used to ultimately build a quantum repeater, and describe recent experiments that have demonstrated the generation of entanglement between a single photon and a quantum dot. In particular, we review experiments that have generated and verified entanglement between the polarization or frequency state of a photon emitted by a single quantum dot, and the spin state of the electron in that quantum dot \cite{de_greve_quantum-dot_2012,gao_observation_2012,schaibley_demonstration_2013}. We also review how tomography can be performed on a spin-photon qubit pair, and describe the result in Ref. \cite{de_greve_complete_2013}, which showed that quantum dots can produce spin-photon entanglement with fidelity in excess of 90\%. We provide a summary of spin-photon entanglement generation results in many different physical systems. We also address briefly some questions surrounding what work needs to happen to proceed from the present state of affairs to a functioning quantum-dot-based quantum repeater.

\section{Quantum Repeaters}

Before we discuss how optically-active quantum dots may be suitable building blocks for constructing a quantum repeater, we would like to provide a general overview of quantum repeaters. We attempt to provide answers to the following questions:

\begin{itemize}
  \item What are quantum repeaters, and why is there substantial interest in building them? \\
	\item What are the technological requirements for building useful quantum repeaters? \\
\end{itemize}

\subsection{Motivation for Quantum Repeaters}

The introduction or motivation sections of many quantum dot papers begin with a brief mention that optically-active quantum dots will be useful for {\it quantum information processing}, or sometimes more specifically, that they will be useful for building quantum repeaters or quantum computers. However, there is a fairly large disconnect between the literature on the engineering of quantum devices (such as quantum dots, but also other systems) and the quantum information theory literature. Furthermore, even the quantum information literature rarely explicitly explains the relationships between the many different low-level protocols and proposals, and how various subsets of them may fit together to enable the construction of high-level quantum technology (such as quantum computers or quantum cryptography).

The main high-level motivation for research in {\it quantum communication} (also known as {\it quantum networks}) is the development of practical long-distance quantum cryptography (which is also more precisely known as {\it quantum key distribution}). As we will explain, quantum repeaters are central to quantum communication research. Quantum cryptography can be implemented in two different ways (non-entanglement-based and entanglement-based), only one of which involves quantum networks and quantum repeaters.

In this section we provide a description of the two main approaches to implementing long-distance quantum cryptography, with a focus on how quantum networks and quantum repeaters are related to this goal. This connection of quantum repeaters to quantum cryptography is the main high-level motivation for building quantum repeaters. However, we begin with a lower-level motivation (a physics-based, rather than application-based motivation) for quantum repeaters, and provide a summary of several important protocols and proposals that are relevant to their design. 

\subsubsection{Physical Motivation for Quantum Repeaters}

A simple description of the purpose of quantum repeaters is that they enable the generation and/or distribution of entangled qubit pairs over long distances\footnote{The connection between long-distance entanglement distribution and quantum cryptography (which is arguably the main current driver behind the quest to build practical quantum repeaters) will be explained shortly.}; without quantum repeaters, it may be impossible to generate entangled qubit pairs at high rates over distances much greater than several hundreds of kilometers.\footnote{This limit is under the assumption of entanglement distribution occurring using photon transmission in optical fibres. However, as we mention briefly later in this chapter, even for free-space transmission in satellite-based schemes, at least one quantum repeater will seemingly be needed to distribute entanglement to the opposite side of the earth.} Throughout this chapter, we will use the term {\it quantum memories} to refer to stationary qubits\footnote{Typically implemented using matter, as opposed to light. We focus on the use of spin qubits stored in quantum dots as quantum memories.} at the network endpoints and in the quantum repeater stations, which is the standard nomenclature in the quantum repeater/communication/networking community. In this language, the goal of quantum repeaters is to enable the entanglement of quantum memories at sites that are spatially separated by large distances. A key part of entanglement distribution protocols (including schemes involving quantum repeaters) is how photons (generically referred to as {\it flying qubits}, and more precisely as {\it photonic qubits}) can be used to mediate the entanglement of distant quantum memories; this is a major theme of this chapter.

One of the fundamental intuitions behind the need for quantum repeaters in quantum communication is the same as the motivation for classical repeaters in classical communication: photon loss in optical fibres (or in free-space) reduces the power of the signal being transmitted \cite{keiser_optical_2003}, and without regeneration of the signal, low-error-rate, high-bandwidth communication becomes impossible. Since it is impossible to clone a single quantum mechanical state \cite{wootters_single_1982,dieks_communication_1982}, quantum repeaters need to use a different method than classical repeaters to transmit quantum information from one node to the next. This is one of the essential goals of {\it entanglement swapping} in quantum repeaters. Entanglement swapping in a repeater network allows an entangled qubit pair to be generated at the endpoints of the network, by linking together qubits that are initially just entangled with those at neighboring nodes. With this resource in place, teleportation \cite{bennett_teleporting_1993} can be used to transmit an arbitrary qubit from one end of the network to the other.\footnote{We note this use of teleportation for the sake of completing the analogy with a classical repeater network, which is used to transmit classical bits from one end of the network to the other. Quantum key distribution, i.e., quantum cryptography, typically does not make use of teleportation.}

Repeaters in classical communication serve another important purpose besides just amplifying the transmitted signal: they perform error correction by recreating high-quality representations of bits from low-quality representations, since distortions caused by transmission through the optical fibre ultimately lead to bit-discrimination errors if left unchecked \cite{morthier_optical_2003}. This purpose of classical repeaters suggests an equivalent function for quantum repeaters in quantum networks: quantum repeaters should correct decoherence in the entangled qubits before the decoherence becomes so severe that it is uncorrectable. The analogy between the error correction task of classical repeaters and quantum repeaters is, however, imperfect, for the following reason. Classical repeaters, for which the primary source of errors that need correcting are those caused by distortions to the signals (electrical or photonic) propagating between repeater sites, can be assumed to have perfect memories and completely error-free local operations on those memories. However, in a quantum network, quantum repeaters not only need to ameliorate the channel-induced decoherence to the flying qubits\footnote{An example of channel-induced decoherence is that caused by uncontrolled birefringence in an optical fibre, when transmitting a polarization-encoded photonic qubit ($\ket{\psi} = \alpha \ket{\text{H}} + \beta \ket{\text{V}}$): this leads to random qubit rotations (resulting in a loss of state fidelity), and polarization mode dispersion (which in turn results in the overlapping of different qubits' temporal wavepackets, and consequently a reduction in entanglement).}, but also the loss in fidelity of the final stationary entangled qubits (quantum memories), which occurs for a myriad of reasons that are unrelated to the channel-induced decoherence of the photonic qubits. One of the dominant reasons is simply the natural decoherence of the physical stationary qubits, characterized by their $T_2$ time. Furthermore, the local quantum operations in each repeater are imperfect, and will cause reductions in fidelity when they are applied. This chapter has a focus on the interface between the stationary qubits and the flying qubits, and as we will see, the fundamental task of generating spin-photon entangled states occurs with remarkably low fidelity in most physical systems. Quantum repeaters need to compensate for all these mechanisms that result in reduced fidelity of the entangled qubit pairs.

One interesting approach to this problem is to use {\it entanglement purification} \cite{bennett_purification_1996,bennett_concentrating_1996}: this is a technique by which two lower-fidelity entangled qubit pairs can be combined (using only local operations) to produce one higher-fidelity entangled qubit pair. The initial proposals \cite{briegel_quantum_1998,dur_quantum_1999} for quantum repeaters analyzed this approach to combating errors. However, this is not the only possibility: a large body of work on error correction for quantum computers has been developed, and much of this work is potentially relevant to quantum repeaters.\footnote{Bennett {\it et al.} \cite{bennett_mixed-state_1996} showed that entanglement purification is deeply connected to quantum error correction; in particular, they showed that entanglement purification with a classical communication channel is equivalent to quantum error correction, so it is not surprising that quantum repeater protocols can in principle make use of either entanglement purification or quantum error correction protocols to distribute high-fidelity states in the presence of noise.} Several contemporary quantum repeater proposals, such as Ref. \cite{fowler_surface_2010}, explicitly call for quantum error correcting codes \cite{devitt_quantum_2013,lidar_quantum_2013} to be used as the mechanism for combating errors in quantum networks, instead of entanglement purification. Hybrid approaches, in which both quantum error correction and entanglement purification are used, have also been proposed \cite{jiang_quantum_2009}.

The connections between the functionality of classical communication repeaters and quantum repeaters are summarized in Table \ref{tab:ClassicalAndQuantumRepeaters}.

\begin{table}
\caption{A summary of the analogues between classical communication repeaters and quantum repeaters.}
\label{tab:ClassicalAndQuantumRepeaters}
\begin{tabular}{p{4cm}p{4cm}p{4cm}}
\hline\noalign{\smallskip}
Problem & Classical Repeater Solution & Quantum Repeater Solution \\
\noalign{\smallskip}\hline\noalign{\smallskip}
Channel-induced Loss & Signal Amplification (via Regeneration) & Entanglement Swapping \\
Channel-induced Distortion & Signal Regeneration & Entanglement Purification / Quantum Error Correction \\
\noalign{\smallskip}\hline\noalign{\smallskip}
\end{tabular}
% footnotes here
\end{table}

\subsubsection{Quantum Key Distribution and Quantum Cryptography}

Over the past two decades, nearly all experimental work on implementing quantum cryptography has focused on schemes derived from one of two sources: the original BB84 protocol \cite{bennett_quantum_1984} (which does not involve entanglement) and the Ekert91 protocol \cite{ekert_quantum_1991,bennett_quantum_1992} (which does rely on entanglement).

The fundamental ideas behind quantum cryptography have been well-explained in many previous review articles and books; we do not repeat them here, but recommend instead References \cite{nielsen_quantum_2011} and \cite{gisin_quantum_2002} as starting points for readers unfamiliar with the BB84 and Ekert91\footnote{We generally refer to the version of the Ekert91 protocol described by Bennett {\it et al.} in Reference \cite{bennett_quantum_1992}.} protocols.

Bennett {\it et al.} \cite{bennett_quantum_1992} showed that the Ekert91 protocol is in some sense equivalent to the BB84 protocol. One might na\"ively conclude that BB84 is a superior choice for practical implementation, since it calls for only a single source of unentangled flying qubits, whereas Ekert91 requires the generation of high-fidelity entangled qubit pairs. However, there is a crucial difference between BB84-based schemes and Ekert91-based schemes that we would like to emphasize here: BB84-based QKD can be achieved over long distances using classical relays that need physical security, whereas Ekert91-based QKD can be achieved over long distances using quantum repeaters that need not be secure. Given that repeaters in a fibre-based network will likely need to be placed somewhere between every $10 \km$ and every $300 \km$, the advantage of not needing trusted, armed guards at every repeater station in order to ensure the integrity of the system is highly non-trivial.

Satellite-based schemes \cite{aspelmeyer_long-distance_2003} largely avoid the need for repeaters, but have their own disadvantages (for example, the ease with which an attacker could perform a denial-of-service attack by simply blocking the free-space path, or by destroying the satellite). Nevertheless, practical satellite-based Ekert91 may well be implemented before fibre-and-quantum-repeater-based Ekert91, due to the extreme difficulty in implementing a practically-relevant quantum repeater. To the extent that satellite-based QKD schemes do use repeaters (for example, for dealing with the lack of a direct free-space path from one side of the earth to the other), our descriptions of classical relays and quantum repeaters, and the potential role of QDs in building these quantum communication technologies, remain relevant. We also note that satellite-based schemes can plausibly implement both BB84-based QKD and Ekert91-based QKD, with many of the same advantages and disadvantages we discuss for fibre-based implementations of either.

\subsubsection{Long-distance Quantum Key Distribution with Classical Relays}

\begin{figure}[t]
\centering
\includegraphics[width=12cm]{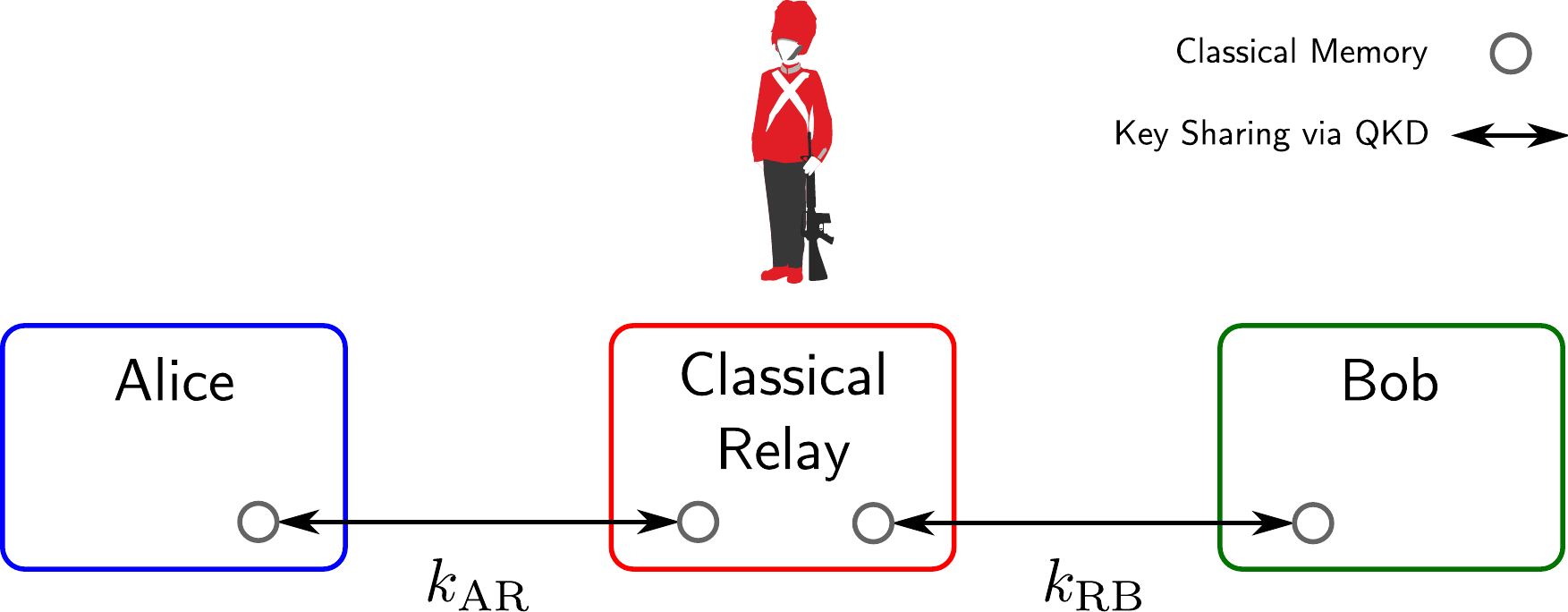}
\caption{Long-distance Quantum Key Distribution using a Classical Relay. If a key $k_\text{AR}$ is shared between Alice and a relay, and another key $k_\text{RB}$ is shared between the relay and Bob, then a secure message can be sent from Alice to Bob, assuming that the relay site is secured. Classical relays are much simpler to build than quantum repeaters, since they only need classical memories. All relays must however be secured, otherwise the privacy of the communications between Alice and Bob cannot be guaranteed.}
\label{fig:QKDRelay}
\end{figure}

Scarani {\it et al.} \cite{scarani_security_2009} provide a comprehensive review of the derivatives of the original BB84 protocol that have been developed over the past 20 years as a result of the challenges in making single-photon sources and in transmitting polarization-encoded qubits over substantial distances without decoherence. In Section VIII.A.5 of Ref. \cite{scarani_security_2009}, they provide a very brief summary of the use of classical relays to extend the distances over which quantum key distribution can work. A recent example of the deployment of such a QKD network is that by a group of companies aiming to build a large network in China, including a link between Beijing and Shanghai \cite{qiu_quantum_2014}.

The idea of a classical relay for QKD is very simple. Suppose we have distant stations for Alice (A), Bob (B), and a relay (R). We begin by having Alice and the Relay share a secret key $k_\text{AR}$ (using, for example, BB84), and having the Relay and Bob share a (different) secret key $k_\text{RB}$. There are now two main options -- Option 1, as described in References \cite{elliott_building_2002,scarani_security_2009}: if Alice wants to send a secure message to Bob, she can encrypt the message using the key $k_\text{AR}$, the Relay can decrypt the message (using the key $k_\text{AR}$), then re-encrypt the message using key $k_\text{RB}$, and send the encrypted message to Bob, who can decrypt the message. In this option, the QKD relay stores both keys and is involved in transmitting the actual message. Option 2, as described in References \cite{elliott_building_2002,dianati_architecture_2007,Peev_SECOQC_2009} and implemented in the Vienna QKD network \cite{Peev_SECOQC_2009}: alternatively the Relay can use the key $k_\text{RB}$ to encrypt a message consisting of the key $k_\text{AR}$ (which is the key Alice holds), and send this message to Bob, who can decrypt it using the key $k_\text{RB}$. Bob thus ends up with the key $k_\text{AR}$, and so a secret key ($k_\text{AR}$) has been distributed between Alice and Bob, via the relay. Alice and Bob can then communicate using this key over whatever classical channel they like. In this option, the QKD relay is only ever used to transfer keys.

As we have noted, the classical relay strategy given here does not depend on the method used to share the private keys between nearest-neighbour stations. Thus long-distance QKD using classical relays in the way we have described here can be performed with BB84-based protocols or Ekert91-based protocols. However, the benefit that Ekert91-based protocols can offer long-distance QKD using insecure repeater stations is only true with we use quantum repeaters; with classical repeater stations, the same repeater station physical security requirement as with BB84-based implementations is imposed.

\subsubsection{Long-distance Quantum Key Distribution using Ekert91 and Quantum Repeaters}
\label{sec:QKDwithQR}

\begin{figure}[t]
\centering
\includegraphics[width=12cm]{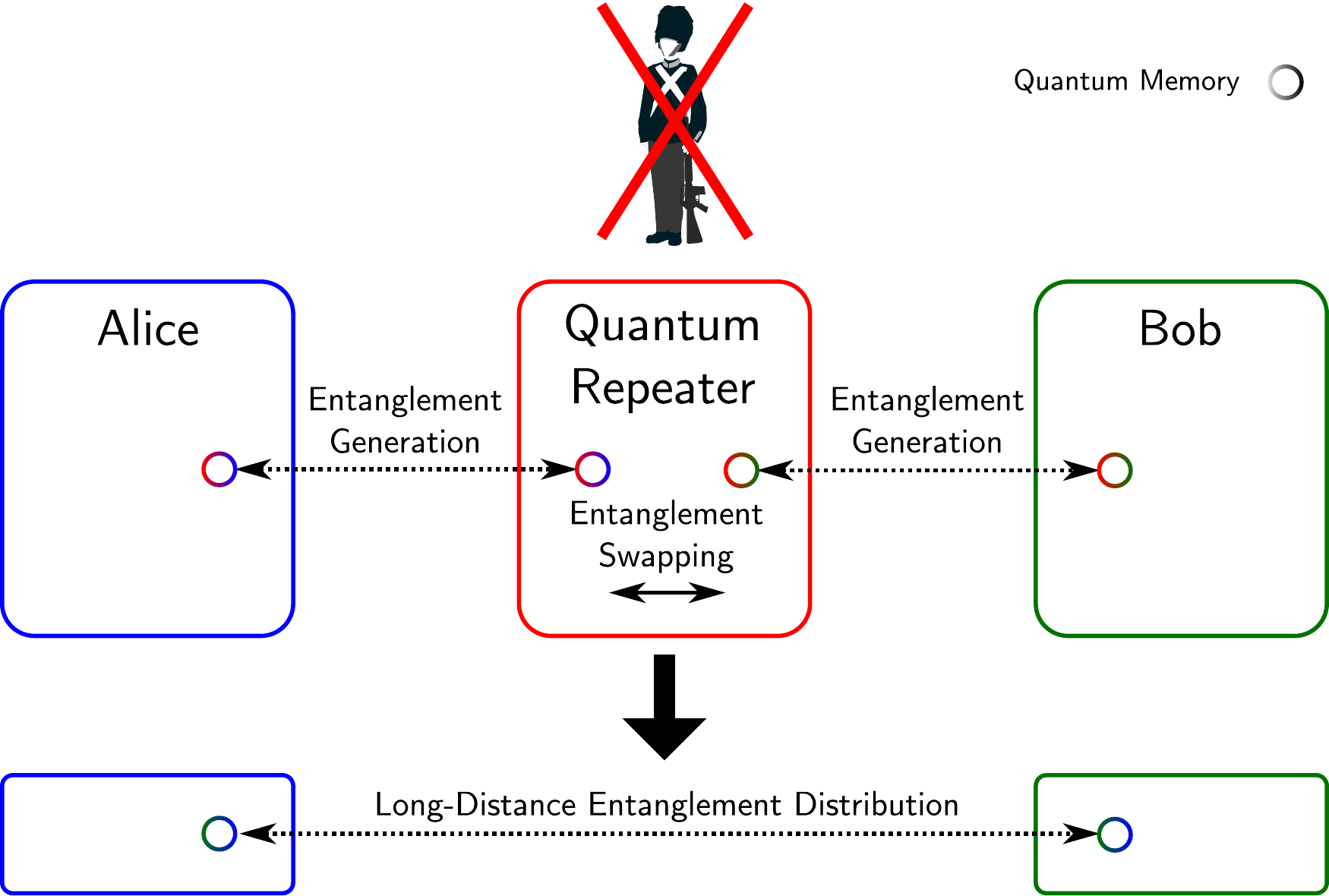}
\caption{Long-distance Entanglement Distribution using a Quantum Repeater. Long-distance quantum key distribution without the need for secured relays is made possible by the use of a quantum repeater network, which can distribute entanglement between the endpoints (Alice and Bob), and an entanglement-based QKD protocol (Ekert91, or derivatives thereof). If an eavesdropper disturbs even one of the quantum repeater stations or links, and attempts to gain information about the key being distributed, this disturbance will be detectable (unlike with the classical relay scheme).}
\label{fig:Repeater}
\end{figure}

A quantum repeater is a device that allows for the distribution of entangled qubits over distances that are beyond the limits imposed by loss and decoherence when considering sending qubits directly from one node (Alice) to another node (Bob). The fundamental advantage that quantum repeaters have over classical relays for extending the range over which QKD is possible is, as we have mentioned, that the quantum repeater nodes need not be physically secure. The main disadvantage that they have is that it appears to be exceptionally difficult to realize practical quantum repeaters.

The Ekert91 protocol \cite{ekert_quantum_1991,bennett_quantum_1992,nielsen_quantum_2011} for QKD between two nodes (Alice and Bob) calls for the generation of an entangled qubit pair where one of the qubits is sent to Alice, and the other is sent to Bob. If we place a quantum repeater between these nodes, the distance between Alice and Bob can be extended. First we need Alice and the Repeater to share an entangled qubit pair, and for Bob and the Repeater to share another entangled qubit pair. Now, at the Repeater node, we perform a measurement of the two qubits in the Bell state basis; the outcome heralds the creation of an entangled Bell qubit pair between the qubits held by Alice and Bob. This procedure is called entanglement swapping, since the qubits that were at Alice and at Bob, which were originally not entangled, become entangled as a result of the local measurement operations that are performed at the Repeater. This is one of the two fundamental operations of a quantum repeater, and was described in 1993 by Bennett {\it et al.} \cite{bennett_teleporting_1993} and by \.{Z}ukowski {\it et al.} \cite{zukowski_event-ready-detectors_1993}. 

The entanglement swapping procedure itself does not require a quantum memory. However, an entanglement-swapping-based quantum repeater must have long-coherence-time quantum memories, otherwise it will be unable to provide a benefit. This is because entanglement swapping only works if entanglement between Alice and the Repeater, and between the Repeater and Bob, exist simultaneously. To illustrate this somewhat explicitly, we consider two scenarios -- one without a repeater, and one with a memoryless repeater, and in both cases an equal total distance ($L$) between Alice and Bob. For concreteness, we assume that entangled-photon-pair sources between the nodes are used as sources of entangled qubits. The transmission probability $p_L$ of a photon propagating through a fibre decays exponentially with the length $L$ of the fibre: $p_L = 10^{\frac{-L\alpha}{10}}$, where $\alpha$ is the attenuation coefficient in dB per unit distance. In the scenario with just Alice and Bob (no repeater), an entangled-photon pair will successfully be shared between Alice and Bob with probability $p_{L/2}^2=p_L$. If the entangled-photon-pair source generates pairs at a rate $R_0$, then the rate of entanglement generation between Alice and Bob will be $R_0 p_L$. In the second scenario, that with Alice, a (memoryless) repeater, and Bob, each separated by a distance $L/2$, entanglement may be generated between Alice and the Repeater with probability $p_{L/2}$, and independently entanglement may be generated between the Repeater and Bob with probability $p_{L/2}$. For entanglement swapping to succeed, both these events must occur simultaneously, which happens with probability $p_{L/2}^2=p_L$, so the rate of entanglement generation between Alice and Bob is $R_0 p_L$. Therefore the entanglement generation rate between Alice and Bob is the same for the case of a direct connection between Alice and Bob, and the case when a memoryless repeater is placed between Alice and Bob.

However, if we endow Alice, Bob, and the Repeater with memory, then we find that using a repeater can increase the rate of entanglement generation over a fixed distance, as we will see with the following toy protocol. The Repeater waits for the photonic qubit from Alice's side to arrive\footnote{The Repeater can also handle the case where the photon from Bob's side arrives first. For simplicitly, we describe here only the case of entanglement being successfully generated between Alice and the Repeater first.}, and stores it in memory.\footnote{We assume here for illustrative purposes that the probability for the repeater to store the flying qubit in memory is unity; in a practical repeater this is an important parameter to optimize, since if it is too small, the use of a repeater will reduce the overall rate of entanglement distribution.} Alice's side is instructed (via a classical channel) to stop generating Bell pairs, and Alice stores her current qubit in memory too.\footnote{Note that we have assumed here that the Repeater can tell if a photon (from Alice or Bob) has arrived. This is in practice difficult to do without disturbing the photon, so repeaters are generally designed to avoid this requirement. We describe more practical proposals in the next section.} A Bell pair is now shared between Alice and the Repeater. Bob's side continues to generate Bell pairs, sending photons to the Repeater. The Repeater waits for a photonic qubit from Bob's side to arrive, and when one does, the Repeater can then perform entanglement swapping between the qubit from Alice's side (stored in the Repeater's memory) and the new qubit from Bob's side, yielding entanglement between Alice and Bob's local qubits. The rate of generation of entanglement between Alice and Bob is\footnote{For the protocol described here, and including the handling of the case where the photon from Bob's side arrives first and is stored, the rate is $\approx (0.67) \cdot R_0 \sqrt{p_L}$.} $\sim R_0 \sqrt{p_L}$. If $p_L \ll 1$, this rate may be substantially faster than $R_0 p_L$ (the rate for the memoryless repeater scheme), so the Repeater has become useful.\footnote{As we cover in more detail in Section \ref{sec:EntanglementGeneration}, $p_L$ is typically very small: loss in the fibre results in $p_L \sim 10^{-2}$ when $L \sim 100 \km$, and coupling and detection losses can easily amount to a further $30 \dB$ of loss. Therefore in basically all practical situations, the rate $(0.67) \cdot R_0 \sqrt{p_L}$ is much larger than $R_0 p_L$. } 

A rudimentary repeater using only entanglement swapping, such as the one described above, may make long-distance entanglement distribution over fibre practical, assuming that the Bell pair generation is perfect, that the quantum memories are perfect, and that the local operations at the repeaters are perfect. Entanglement purification \cite{bennett_concentrating_1996,bennett_purification_1996,deutsch_quantum_1996} allows some of these assumptions to be relaxed. As we have mentioned already, entanglement purification refers to a class of procedures that each use a set of lower-fidelity entangled qubit pairs to produce a smaller number of higher-fidelity entangled qubit pairs, provided that the fidelity of the initial qubit pairs is above a certain threshold. Entanglement purification provides a clever solution to deal with the imperfections of a real system, since the effect of all imperfections is just the degradation of the fidelity of the entangled qubit pairs. Some of the early quantum repeater proposals \cite{briegel_quantum_1998,dur_quantum_1999} analyzed how one may perform long-distance entanglement distribution using quantum repeaters (incorporating both entanglement swapping and entanglement purification) that have faulty local operations, and found that error rates of $\sim 1 \%$ for local one- two-qubit gates and measurement may be tolerated (i.e., the system may still be able to distribute high-fidelity entangled pairs, even when the local operations in the repeaters are imperfect).

Unfortunately, achieving the assumed fidelities and operation error probabilities in experimental systems is very challenging. Furthermore, it is unreasonable to assume that physical stationary qubits will be arbitrarily long-lived, and in the case of spins in quantum dots, it is unlikely that $T_2$ times beyond several milliseconds will be achievable \cite{kroutvar_optically_2004,press_ultrafast_2010}, even with substantial materials and device engineering effort. Fortunately, it is in principle possible to make an arbitrarily long-lived logical quantum memory by using quantum error correction \cite{lidar_quantum_2013}, provided enough physical qubits are available, and sufficiently high-fidelity local operations can be performed on them. Building quantum repeaters using a fault-tolerant error correcting scheme also allows for the construction of logical local operations with fidelities that are much higher than the fidelities of the native operations on physical qubits. 

More recent theoretical work on quantum repeaters has also considered how to perform the task of entanglement purification (which is effectively that of correcting errors in the distributed Bell pairs) using other methods based on fault-tolerant quantum error correction, such as Calderbank-Shor-Steane codes \cite{jiang_quantum_2009}, the surface code \cite{fowler_surface_2010,fowler_surface_2012}, and topologically-protected cluster states \cite{li_long_2013}. These approaches may also have advantages over entanglement-purification-based quantum repeaters \cite{briegel_quantum_1998,dur_quantum_1999} in the reduced classical communication required for operation, which is predicted to have dramatic effects on performance \cite{jiang_quantum_2009,fowler_surface_2010}.

The high-level architectural studies of quantum repeaters are currently far-removed from practical experimental realities, and we will not go into further detail about them in this chapter. However, one important overall point for us to emphasize is that these state-of-the-art proposals for quantum repeaters essentially call for the implementation of quantum repeaters as small\footnote{Fowler {\it et al.} \cite{fowler_surface_2010} predict that their scheme will be able to distribute entangled pairs from one side of the earth to the other at a MHz rate if the endpoints are connected by $\sim 10^4$ repeaters, each containing $\sim 10^3$ physical qubits, provided that initial entangled pair fidelities are $\gtrsim 0.96$, and quantum gates that can operate on nanosecond timescales are available.} fault-tolerant quantum computers that are also equipped with photonic interfaces. The task of constructing practical quantum repeaters thus appears to be at least as difficult, if not more difficult than, building a practical fault-tolerant gate-model quantum computer.

\subsection{Design of Quantum Repeaters}

As we have explained in the previous section (Section \ref{sec:QKDwithQR}), quantum repeaters need to incorporate quantum memory. One approach is to directly store photonic qubits, for example using a cavity.\footnote{One can imagine storing a photonic qubit in a ring cavity, but sufficiently low-loss cavities are not available in practice. For example, to store a photon in a fibre loop for $1 \ms$ requires the photon to propagate through $(1 \ms \cdot \frac{c}{n_\text{core}}) \approx 185 \km$ of fibre, which would result, at best \cite{corning_pi1470:_2014}, in absorption of the photon (and consequently complete loss of the qubit) in $\approx 99.9 \%$ of attempts to store the qubit. To achieve on-demand photon extraction for variable storage times, a slightly more sophisticated cavity scheme is needed (such as that described in Ref. \cite{pittman_cyclical_2002}), which typically introduces even more loss.} The alternative, which we focus on, is to introduce quantum memories based on matter, and an interface between these quantum memories and photons (both for incoming and outgoing photons).

\subsubsection{Heralded Entanglement Generation}
\label{sec:EntanglementGeneration}

In 2001, Duan {\it et al.} \cite{duan_long-distance_2001} introduced a protocol (known as the {\it DLCZ} scheme) for entangling two remote atomic-ensemble-based quantum memories, using photons, and in such a way that successful entanglement is heralded\footnote{By {\it heralded}, we mean that although the protocol for generating entanglement does not succeed every time it is attempted (and indeed may have an extremely low probability of success), the protocol intrinsically provides a signal that lets the experimenter know when the protocol was successful.}. The DLCZ protocol is a member of a class of heralded protocols that can be used to entangle distant quantum memories provided that it is possible to generate an entangled state between each quantum memory and a photonic qubit.

Another protocol from this class is the Simon-Irvine protocol \cite{simon_robust_2003}. The treatment of it that we give here follows closely the formulation given by Moehring {\it et al.} \cite{moehring_entanglement_2007}.\footnote{While we have chosen to focus on one particular heralded entanglement generation protocol in this chapter, we don't wish to give the impression that this is the only protocol that can possibly be used to entangle spins in remote quantum dots. Our discussion of a variant of the Simon-Irvine protocol is motivated by the fact that it is applicable to quantum dots, and has been successfully demonstrated with single ions \cite{moehring_entanglement_2007}. However, many other protocols exist that may plausibly be used to entangle remote quantum dot spin qubits, and may ultimately prove to be superior. Our discussion is meant merely to provide intuition for how one popular subset of protocols (those involving spin-photon interfaces and single-photons) works.} Assume that we have two remote quantum memories, Alice (A) and Bob (B), and each memory can be described as a single qubit: Alice has memory basis states $\left\{ \ket{\uparrow}_\text{A} , \ket{\downarrow}_\text{A} \right\}$, and Bob has memory basis states $\left\{ \ket{\uparrow}_\text{B} ,  \ket{\downarrow}_\text{B} \right\}$. Let's suppose that each memory can be entangled with a polarization-encoded photonic qubit, i.e., each quantum memory has associated with it a single photon whose polarization state we use to represent a qubit. We will label the basis states of the photonic qubit for Alice as $\left\{ \ket{\text{H}}_\text{A} , \ket{\text{V}}_\text{A} \right\}$, and for Bob as $\left\{ \ket{\text{H}}_\text{B} ,  \ket{\text{V}}_\text{B} \right\}$.

Suppose that both Alice and Bob can, through some as-yet-undescribed method, produce the following spin-photon entangled states:

\begin{eqnarray}
  \ket{\psi}_\text{A} &=& \frac{1}{\sqrt{2}} \left( \ket{\uparrow}_\text{A} \otimes \ket{\text{H}}_\text{A} - \ket{\downarrow}_\text{A} \otimes \ket{\text{V}}_\text{A} \right) \\
	\ket{\psi}_\text{B} &=& \frac{1}{\sqrt{2}} \left( \ket{\uparrow}_\text{B} \otimes \ket{\text{H}}_\text{B} - \ket{\downarrow}_\text{B} \otimes \ket{\text{V}}_\text{B} \right)
\end{eqnarray}

In this case, Alice has a quantum memory and a photon that is entangled with it, and similarly Bob has a quantum memory, and a photon that is entangled with it. The key idea of the protocol is that we can perform a simple operation that will perform entanglement swapping on the photons from Alice and Bob, such that when the entanglement swapping operation has been completed, the two quantum memories of Alice and Bob will be entangled, even though they never directly interacted with each other.

\begin{figure}[t]
\centering
\includegraphics[width=12cm]{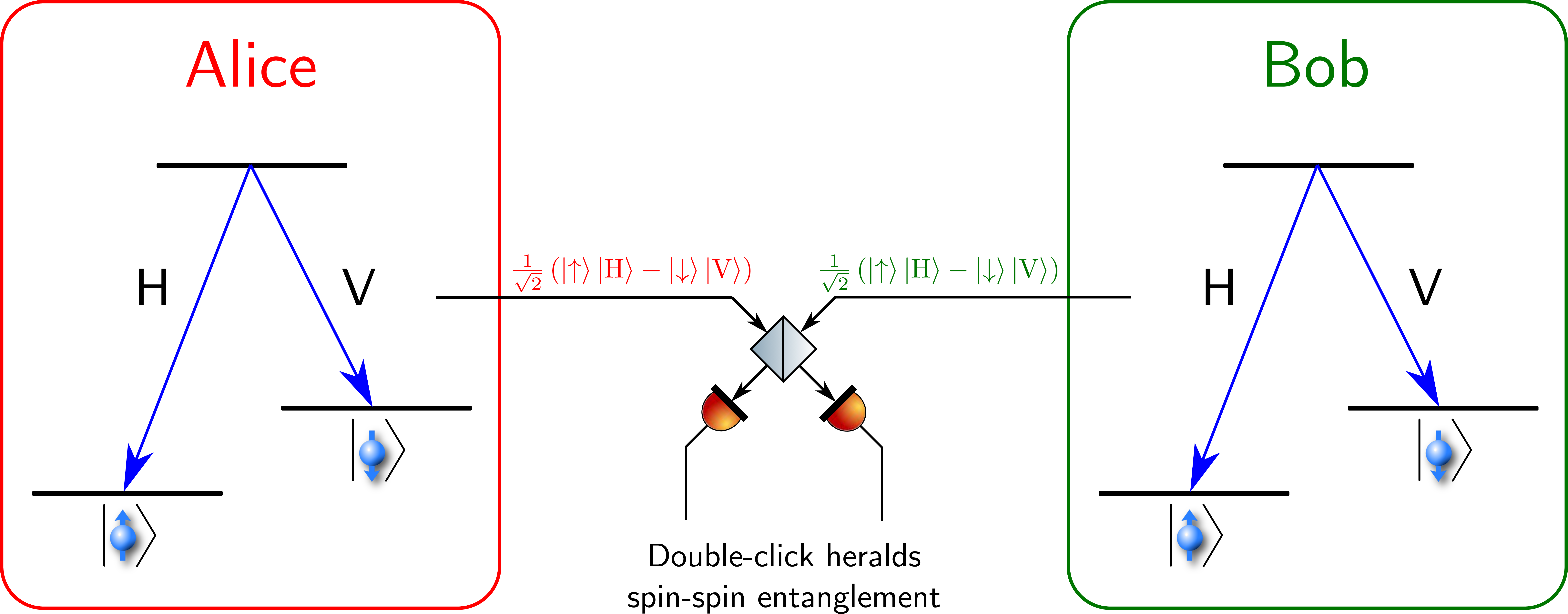}
\caption{A Protocol for Entangling Remote Quantum Memories. A quantum memory (represented here by a single spin) located with Alice can be entangled with a quantum memory located with Bob if both quantum memories can emit photons that are entangled with their respective spins. The photons from Alice and Bob are interfered on a beam splitter, which has detectors on both output ports. If both detectors detect a photon, then entanglement between Alice's and Bob's memories has been generated. In particular, the following maximally-entangled state of Alice's and Bob's memories is heralded: $\ket{\Psi^-}_\text{memories} = \frac{1}{\sqrt{2}} \left( \ket{\uparrow}_\text{A} \ket{\downarrow}_\text{B} - \ket{\downarrow}_\text{A} \ket{\uparrow}_\text{B} \right)$.}
\label{fig:EntanglementGenerationProtocol}
\end{figure}

Figure \ref{fig:EntanglementGenerationProtocol} illustrates this entanglement generation protocol. The photon from Alice and the photon from Bob are mixed on a non-polarizing 50/50 beamsplitter, and each output port of the beamsplitter is monitored by a single-photon detector (which produces a click if a photon is present in the mode, and otherwise does not). The state of the system before the beamsplitter is:

\begin{eqnarray}
  \ket{\psi}_\text{system} &=& \ket{\psi}_\text{A} \otimes \ket{\psi}_\text{B} \\
	&=& \frac{1}{2} \left[ \left( \ket{\uparrow}_\text{A} \ket{\text{H}}_\text{A} - \ket{\downarrow}_\text{A} \ket{\text{V}}_\text{A} \right) \otimes \left( \ket{\uparrow}_\text{B} \ket{\text{H}}_\text{B} - \ket{\downarrow}_\text{B} \ket{\text{V}}_\text{B} \right) \right] \\
	&=& \frac{1}{2} \Bigl[ \ket{\Phi^+}_\text{memories} \ket{\Phi^+}_\text{photons} + \ket{\Phi^-}_\text{memories} \ket{\Phi^-}_\text{photons} - \nonumber\\
	&&\quad \ket{\Psi^+}_\text{memories} \ket{\Psi^+}_\text{photons} - \ket{\Psi^-}_\text{memories} \ket{\Psi^-}_\text{photons} \Bigr]
\end{eqnarray}

As given by Moehring {\it et al.} \cite{moehring_entanglement_2007}, this rewriting of the system state in terms of states of the memories and of the photons allows us to easily interpret the outcomes of such a setup. Here $\ket{\Phi^\pm}_\text{memories} = \frac{1}{\sqrt{2}} \left( \ket{\uparrow}_\text{A} \ket{\uparrow}_\text{B} \pm  \ket{\downarrow}_\text{A} \ket{\downarrow}_\text{B} \right)$, and $\ket{\Psi^\pm}_\text{memories} = \frac{1}{\sqrt{2}} \left( \ket{\uparrow}_\text{A} \ket{\downarrow}_\text{B} \pm  \ket{\downarrow}_\text{A} \ket{\uparrow}_\text{B} \right)$. Identical photons impinging on a beamsplitter give rise to the Hong-Ou-Mandel effect \cite{hong_measurement_1987}: they will bunch into the same output port. For photons that are indistinguishable in all but their polarization, the net effect gives rise to a situation where only a fully antisymmetric two-photon state\footnote{More precisely, the quantum state describing the polarization degree of freedom of each photon should be antisymmetric.} impinging on the beamsplitter in this experimental setup (Figure \ref{fig:EntanglementGenerationProtocol}) can result in both detectors clicking at the same time. Any symmetric two-photon input state leads to photon bunching, where both photons exit out of a single port, resulting in (at most) only one of the detectors clicking in the relevant time window.\footnote{This assumes the absence of detector dark counts.} Of the four two-photon states $\ket{\Phi^\pm}_\text{photons},\ket{\Psi^\pm}_\text{photons}$, only $\ket{\Psi^-}_\text{photons} = \frac{1}{\sqrt{2}} \left( \ket{\text{H}}_\text{A} \ket{\text{V}}_\text{B} - \ket{\text{V}}_\text{A} \ket{\text{H}}_\text{B} \right)$ is antisymmetric. Therefore if both single-photon detectors after the beamsplitter click, we have measured the photonic part of the system state to be $\ket{\Psi^-}_\text{photons}$, and therefore the memories are projected to be in the state $\ket{\Psi^-}_\text{memories}$. Therefore a double-click event heralds the generation of entanglement between the quantum memories at Alice and Bob's nodes.\footnote{If the single-photon detectors are replaced with number-resolving detectors, then all four memory Bell states can be heralded. If four single-photon detectors (and two polarizers) are available, then both $\ket{\Psi^+}_\text{memories}$ and $\ket{\Psi^-}_\text{memories}$ can be heralded.} 

{\bf Heralding and Experimental Errors}

The use of the double-click event to herald success is very important. There are many ways for such an experiment to result in only one of the detectors clicking\footnote{Some examples are: loss of a photon during outcoupling from the quantum memory system; loss during propagation; failure of a detector to click even though the photon arrived, due to non-unity quantum efficiency of the detector.}. However, so long as detector dark counts are sufficiently low, there can be a high probability that if both detectors click that this was because the photonic state really was $\ket{\Psi^-}_\text{photons}$, so the memories are in the entangled state $\ket{\Psi^-}_\text{memories}$.

Besides imperfections in the detectors (leading to dark counts), there is another way in which this protocol can falsely indicate that $\ket{\Psi^-}_\text{memories}$ has been generated, when in fact it has not. If the quantum memories produce, with non-zero probability, more than one photon within the time window being considered for detector clicks, then the experimentalist may measure two clicks, but have the memories not actually be in the state $\ket{\Psi^-}_\text{memories}$, i.e., the heralded state will not be the target state. This is undesirable. Therefore the second-order correlation function, $g^{(2)}(\tau)$, and in particular, the value $g^{(2)}(\tau = 0)$, is an important parameter for determining the suitability of a quantum-memory--photon interface for use in a quantum repeater. Ideally $g^{(2)}(0) = 0$, and the larger it is, the greater will be the percentage of heralding events that incorrectly indicate that the target entangled state has been generated.

{\bf Impact of Photon Loss on the Effectiveness of Entanglement Distribution}

Protocols that rely on a double-click event (in the way we have described) to herald the generation of entanglement are sensitive to loss. For an attempt at the heralded generation of entanglement between quantum memories to succeed, a photon from Alice must arrive at a detector (and be detected by it), and a photon from Bob must arrive and be detected. Therefore the probabilities $p_\text{A}$, $p_\text{B}$ of photons from Alice and Bob being detected determine the probability $p_\text{success}$ of successful heralded entanglement generation between Alice and Bob as $p_\text{success} = p_\text{A} p_\text{B}$. We have mentioned several ways in which photon loss may occur, but here let's assume that we have the nearly ideal scenario that the only loss is due to absorption in an optical fibre. We now briefly analyze how this photon loss affects the system performance as a function of distance between Alice and Bob.

One of the lowest-loss optical fibres currently available has an attenuation of $\alpha \approx 0.17 \text{~dB/km}$ \cite{corning_pi1470:_2014}, when transmitting photons with wavelength $\sim 1550 \nm$.\footnote{What we outline in this section is a best-case scenario for the loss, since we assume that the photons are in the lowest loss band (covering approximately the range $\SIrange{1525}{1575}{\nm}$ \cite{corning_pi1470:_2014}). Note that the vast majority of current quantum technology experiments occur with systems that emit photons at wavelengths that experience dramatically higher attenuation. For example, the attenuation coefficient for $850 \nm$ photons is typically $\alpha \sim 3.5 \text{~dB/km}$. See Table \ref{tab:SpinPhotonComparison} for a few examples. This strongly motivates work to either engineer quantum systems that natively emit $\sim 1550 \nm$ photons, or to build nonlinear optical systems \cite{zaske_visible--telecom_2012,pelc_downconversion_2012,de_greve_quantum-dot_2012} that can convert light at high-loss wavelengths to wavelengths that have low loss in fibres.} Let's suppose that Alice and Bob's memories can emit photons entangled with them at a rate of $R_0 = 1 \MHz$ (in general, the rate $R_0$ cannot be faster than the inverse of the lifetime of the optically excited state in the quantum memory, which we refer to as the {\it spontaneous emission time}). With perfect photon collection and perfect detectors, the entangled memory generation rate would be $R = 1 \MHz$, in the absence of photon loss. Now let's consider the impact of photon loss in the fibre. Let's suppose that Alice and Bob are a distance $2L$ apart, that the beamsplitter and detectors are located at the midpoint of Alice and Bob, and thus that both memories emit photons into fibres of length $L$. The photon loss in the fibre results in a reduction of photon transmission probabilities: $p_\text{A} = p_\text{B} = 10^{\frac{-L\alpha}{10}}$. Thus the entanglement generation rate $R$ is:

\begin{eqnarray}
  R &=& R_0 p_\text{A} p_\text{B} = R_0 p_\text{success} \\
    &=& R_0 \left( 10^{\frac{-L\alpha}{10}} \right)^2
\end{eqnarray}

For $L = 10 \km$, the loss in each fiber is $1.7 \text{~dB}$, so $p_\text{A} = p_\text{B} \approx 0.68$, thus $p_\text{success} \approx 0.46$, so the entanglement rate $R$ drops to $R \approx 460 \kHz$. If $L = 100 \km$, then $p_\text{success} \approx 4.0 \times 10^{-4}$, so $R \approx 400 \Hz$. If $L = 200 \km$, then $R \approx 0.1585 \Hz$. And if $L = 300 \km$, then $R \approx 6.3 \times 10^{-5} \Hz$; note that this implies the successful generation of an entangled pair only once every $\frac{1}{R} \approx 4.4 \hr$.\footnote{We would like a high rate of entangled-pair generation in general (for example, to facilitate a high generation rate of distributed keys in QKD applications), so naturally we seek to maximize the success probability $p_\text{success}$. However, there is also a crucial limit to how low the heralded success rate can be before the entanglement distribution stops working at all: the rate of coincident arrivals of photons at the detectors needs to be higher than the dark count rate of the detectors (in the appropriate time windows). If the coincident arrival rate is not much higher, then a significant portion of the double-click events will be a result of dark counts, not actual two-photon detections, and these falsely-heralded events will result in a reduction in the fidelity of the target entangled state to below the fidelity threshold for error correction or purification to function. As we will discuss shortly, unheralded entanglement generation protocols lead to mixed states, which is undesirable. False heralding events (for example, due to detector dark counts) also result in mixed states being produced, but with sufficiently low dark-count rates versus heralding success rates, even imperfect heralded schemes are tolerant to photon loss such that rather high fidelity mixed states can be produced (ones that, for example, can still violate Bell's inequality).} It is clear from this simple calculation why quantum repeaters are necessary to generate entanglement over fibre for distances of $\gg 100 \km$. When one considers the other losses in the system, estimates for the distances over which entanglement distribution can be performed through fibre without quantum repeaters are even smaller. 

{\bf The Importance of Heralding for Entanglement Distribution in a Quantum Network}

\nopagebreak
Heralding is important, for at least two reasons: 1.) non-heralded entanglement protocols result in a mixed state $\rho = q \ket{\Psi}\bra{\Psi} + \sum_i r_i \ket{\phi}_i \bra{\phi}_i $, where $\ket{\Psi}$ is the desired (target) entangled state (for example, one of the Bell states), $\ket{\phi}_i$ are other states, and $r_i$ are the probabilities\footnote{The $r_i$ should satisfy the relation $\sum_i r_i = (1-q)$.} of the system ending in one of these states. The state $\rho$ will not violate Bell's inequality, and will generally fail to serve as a useful quantum information resource, if the success probability $q$ is not sufficiently high (as opposed to a heralded scheme, where $q$ can be arbitrarily low, and you can still measure Bell inequality violations provided that you rerun the experiment of generating and measuring the state sufficiently many times that you do actually obtain a set of successfully heralded states). In unheralded schemes, reductions in $q$ directly reduce the fidelity of the output state.\footnote{If all the other states $\ket{\phi}_i$ are not very ``different'' from the target state $\ket{\Psi}$, i.e., $\braket{\Psi}{\phi}_i \sim 1$, then the reduction in fidelity from measuring the mixed state $\rho$, as opposed to the heralded ensemble of target states, will not be severe. However, in many situations, there will be some states $\ket{\phi}_i$ that are nearly orthogonal to $\ket{\Psi}$, and have high probabilities $r_i$ of being generated, and this will dramatically decrease the measured fidelity.} 2.) As we explained in Section \ref{sec:QKDwithQR}, quantum repeaters only confer an advantage if they have quantum memories, since the memory allows for one link to stop trying to generate entanglement after it succeeds. However, if there is no heralding mechanism, there is no way to know when to tell a particular link to stop trying to generate entanglement because it has succeeded!

We can consider the impact on the performance of a quantum network where entanglement generation between nodes is performed with heralding or without heralding in the following way. Suppose we have a network with $N$ nodes (Alice, Bob, and $N-2$ repeater nodes), and that the entanglement generation between adjacent nodes succeeds, on each attempt, with probability $p_\text{success}$, and assume that attempts can be made at a rate $R_0$. With a heralded entanglement generation protocol, the overall rate of entanglement generation between Alice and Bob will scale roughly as\footnote{To illustrate our point, we assume here a simple entanglement-swapping-based approach to distributing entanglement, in which the adjacent nodes each attempt to become entangled with their immediate neighbours (stopping once they have succeeded), and where the protocol is reset once every pair of adjacent nodes shares an entangled qubit pair. This yields an unbroken chain of entanglement that can be converted, via entanglement swapping, to an entangled qubit pair being shared between Alice and Bob. Once an entangled qubit pair is shared between Alice and Bob, we assume it is used, and protocol begins all over again.} $R_0 p_\text{success} / {\log(N)}$, where we note that there is only a very weak (inverse logarithmic) dependence on the number of nodes. However, if the entanglement generation protocol is unheralded, then the rate is dramatically reduced: it will scale as $R_0 p_\text{success}^N$. Note that for even very small numbers of repeaters (e.g., 10), the rate will become unusably small for realistic single-hop success probabilities ($p_\text{success} \ll 1$).

\subsubsection{Constraints on Entanglement Distribution and on Quantum Repeater Design from Finite Quantum Memory Coherence Time}
\label{sec:T2timeConstraints}

In this section, we briefly outline how the coherence times of the quantum memories used impact both simple entanglement distribution experiments, and the design of quantum repeaters for more advanced experiments that incorporate entanglement swapping and purification and/or quantum error correction.

{\bf Constraints on Simple Entanglement Distribution from Finite Quantum Memory Coherence Time}

The current state-of-the-art in experimental demonstrations of entanglement distribution between quantum memories is the generation of entangled states between two quantum memories that are spatially separated by several meters, either through free-space photon propagation, or through optical fibre. This has been achieved with quantum memories implemented in a variety of physical systems, including single $^{171}\text{Yb}^+$ ions \cite{moehring_entanglement_2007}, single $^{87}\text{Rb}$ atoms \cite{hofmann_heralded_2012,ritter_elementary_2012}, ensembles of Cs atoms \cite{chou_measurement-induced_2005}, and with NV centers in diamond \cite{bernien_heralded_2013}. Entanglement between spins in distant quantum dots has not yet been demonstrated. 

Before quantum repeaters using error correction (such as in Refs. \cite{jiang_quantum_2009,fowler_surface_2010,li_long_2013}) become practical, prototype repeaters using no error correction are likely to be tested. In these demonstrations, the coherence time of the quantum memory qubits is a crucial parameter. In the case of qubits formed from spins in quantum dots, $T_2 \gg T_1$, so the $T_2$ time provides the limit on how long the spin can store a qubit.\footnote{Since the longitudinal relaxation time $T_1$ adheres to the relation $2 T_1 > T_2$ (under the assumption that the noise is isotropic with respect to the different qubit axes; this is a good assumption in most systems for physically-relevant noise sources), the $T_1$ time is generally not the limiting timescale. $T_2$, the coherence time (or transverse relaxation time), is generally what defines the useful lifetime of a qubit.} 

Suppose that for the purposes of demonstrating quantum repeater functionality with just two end nodes (Alice and Bob) and a single repeater (endowed only with two quantum memories, and entanglement swapping capability), one uses the following simple protocol. The protocol repeatedly attempts to form entanglement between Alice and the Repeater, and between the Repeater and Bob, and pauses entanglement generation over one of those hops when entanglement is successfully generated over it. In this protocol, the $T_2$ times of the memories at Alice, Bob, and the Repeater should be larger than the time required for the photons to propagate to the midpoint heralding apparatus, in addition to the time required to classically communicate that entanglement generation between Alice and the Repeater (for example) was successful (this will be at least the time required for light to travel half the distance between Alice and the Repeater).\footnote{Jones {\it et al.} \cite{jones_high-speed_2013} introduced a scheme whereby the heralding is performed at the repeater sites (as opposed to at locations midway between the repeaters), and failed attempts can be reattempted without waiting for a delayed classical signal. Even in this protocol though, when a node measures a heralding success, it still has to wait for a classical signal from the adjacent node.} Thus we obtain the limit $T_2 > (\frac{L}{2c} + \frac{L}{2c}) = \frac{L}{c}$. The $T_2$ times should also be longer than the time required to perform the entanglement swapping operation on the quantum memories in the repeater, i.e., $T_2$ should be longer than the one-qubit-gate, two-qubit-gate, and measurement times. For any long distance $L$, the limit from the photon propagation time ($T_2 > \frac{L}{c}$) will be the more stringent one, but for prototype demonstrations (e.g., $L = 10 \meter$), the limit from the local operation times may be more relevant. However, the use of memory is not particularly helpful in improving the rate of generation of entanglement between Alice and Bob if the memories cannot store the qubits for substantially longer than it takes to attempt generating entanglement over a single hop (e.g., between the Repeater and Bob). To demonstrate a substantial benefit from the use of the repeater in distributing entanglement between Alice and Bob, it is necessary for $T_2$ to be at least on the order of the average time it takes for heralded generation of entanglement over a single hop to succeed.\footnote{If the time taken to make a single attempt at generating entanglement over a single hop, set by the distance between the nodes, is $T_\text{rep}$, and the probability of success is $p_\text{success}$, then we want $T_2 \gtrsim T_\text{rep} /{p_\text{success}}$. } Note that meeting this criterion with current technology is not trivial: even for very short distances (on the order of meters), the $T_2$ time will likely need to be seconds\footnote{The repetition time $T_\text{rep}$ will be determined by how quickly the heralding signal can be processed by a classical feedback circuit. Let's assume $T_\text{rep} \sim 1 \us$. Over short distances, $p_\text{success}$ will be dominated by losses other than those from absorption in the fibre; e.g., coupling losses. A reasonable value to assume for quantum dots is $p_\text{success} \sim 10^{-6}$. Thus $T_2 > T_\text{rep} /{p_\text{success}} \sim 1 \sec$.}. If one wants to add additional repeaters in such a demonstration experiment, then the $T_2$ time needs to be increased accordingly.

{\bf Constraints on Quantum Repeater Design from Finite Quantum Memory Coherence Time}

There are many possible designs for a fault-tolerant quantum repeater, and we don't aim to provide comprehensive coverage of them in this chapter. However, given the rather dire predictions in the previous section for what quantum memory coherence times are necessary in order to gain an advantage from using quantum repeaters, we would like to now provide a very brief summary of how the required physical qubit $T_2$ time may be dramatically reduced to values that are more conceivable for quantum dot spin qubits.

For a long-distance quantum network with many hops, without the use of error correction, the physical qubit $T_2$ time may need to be many hundreds, or possibly even thousands, of seconds, in order for the network to sustain a reasonable rate of high-fidelity entanglement generation. Very few physical qubit implementations offer such $T_2$ times, and certainly not quantum dot spin qubits, which seem unlikely to surpass $\SIrange{10}{100}{\ms}$ \cite{kroutvar_optically_2004,press_ultrafast_2010}. 

As we have mentioned before, the general plan in the quantum repeater community for alleviating this problem is to not use physical qubits directly as quantum memories, but rather to implement some form of quantum error correction scheme, in which many physical qubits encode a single logical qubit. Then, so long as local gate operations are sufficiently fast and of sufficiently high fidelity, a logical qubit can be constructed to have an arbitrarily long coherence time (where the ratio of physical qubits required to implement a single logical qubit increases as the desired coherence time increases). For example, the surface code may be able to suitably protect qubits that have $T_2 \sim 100 \us$, provided that nearest-neighbour single-qubit gates, two-qubit gates, and measurement, are available on nanosecond timescales, and with an encoding where $\sim 1000$ physical qubits are used to encode a single logical qubit (quantum memory) \cite{fowler_surface_2010,jones_layered_2012}. 

The prospect of, for each repeater, essentially implementing a fault-tolerant universal quantum computer with thousands of physical qubits, is daunting. There is much work underway to try to find repeater designs that may be more realistically implemented in the near- to medium-term, but currently all proposals require either error rates, or scalability, or both, that are far out of reach of current technology. For a review of many of the leading contemporary proposals, we recommended Ref. \cite{van_meter_quantum_2014}.

\section{Quantum Dots as Building Blocks for Quantum Repeaters}

We have until now described in a fairly abstract way the necessary features and functions of a quantum repeater. There are many physical systems that are currently being considered as candidates for implementing quantum repeaters. Some of them offer the advantage of high native (non-error-corrected) fidelities, which may allow small-scale demonstrations of quantum repeater functionality via entanglement swapping, but which suffer from poor prospects for scalability, which likely will prevent their adoption in building large-scale quantum repeater systems. 

Optically-active charged quantum dots are an appealing candidate physical system for building a high-bandwidth quantum network; one aspect of their appeal is that quantum dot development can leverage progress in commercial semiconductor technology. Schneider {\it et al.} \cite{schneider_lithographic_2008}, Maier {\it et al.} \cite{maier_site-controlled_2014}, and others have succeeded in growing regular 2D arrays of single InAs quantum dots. Jones {\it et al.} \cite{jones_layered_2012} discussed the prospects for designing a large-scale quantum computer that can integrate $>10^8$ quantum dots (each one implementing a single physical qubit) on a single $\sim 4~\text{cm}^2$ chip; one can imagine a very similar design being relevant for a quantum repeater node, except that an additional outcoupling of each photonic interface quantum dot to fibre would need to be implemented. Unfortunately the goal of realizing a $10^8$-physical-qubit quantum computer using quantum dots is still sufficiently divorced from experimental reality that it's not even possible to predict with any certainty when or if it will be possible to realize such a machine. However, if a many-physical-qubit machine can be realized, it is possible that a high-bandwidth repeater system could be implemented despite the large overhead imposed by the use of an error correction code such as the surface code.

Besides the requirement for many physical qubits if one implements a quantum repeater using a large-overhead error correcting code, there is another advantage to having repeater nodes with many quantum memories and photonic interfaces per node: it should be possible to attempt to generate entanglement between memories in adjacent nodes via many channels in parallel, and this will allow for much higher rates of entanglement generation than if only a few parallel channels (or just a single channel) are used. 

Arguably the major fundamental disadvantage of using quantum dots to implement a quantum repeater is the need for the semiconductor sample to be cooled to liquid helium temperatures. At temperatures significantly above $10 \kelvin$, the optical properties of quantum dots degrade dramatically. Many quantum dot spin qubit experiments also currently use superconducting magnets (which are kept at $T \lesssim 4.2 \kelvin$), although it is conceivable that lower magnetic fields (achievable using non-superconducting magnets) may be sufficient.\footnote{The main reason that large magnetic fields (up to $B \sim 6 \tesla$) are currently used is to ensure high-fidelity initialization and readout, when these two operations are performed using optical pumping \cite{de_greve_ultrafast_2013}. However, if high-fidelity, single-shot, quantum nondemolition readout is realized (which is currently thought to be required for any gate-model large-scale quantum computing system \cite{jones_layered_2012}), then it is quite plausible that only small magnetic fields ($B \ll 1 \tesla$) may be required, since there exist proposals for single-shot readout of spins in quantum dots that do not require large magnetic fields \cite{puri_single-shot_2014,puri_optical_2014}.} The use of cryogenic equipment at every repeater station is in principle feasible. However, given the cost of such equipment, there is a strong motivation to find physical systems that offer the advantages of quantum dots, but with the possibility of room-temperature ($T \sim 300 \kelvin$) operation. 

One common standard for coarsely evaluating a candidate physical realization of qubits for implementing a quantum repeater is the set of ``Five (Plus Two)'' DiVincenzo criteria \cite{divincenzo_physical_2000}. The first five DiVincenzo criteria were initially intended for helping to evaluate the suitability of physical qubits for implementing quantum computers. However, as we have covered, most designs for fault-tolerant quantum repeaters call for the creation of machines that are very similar to general-purpose quantum computers, so the DiVincenzo criteria are also relevant when evaluating technology for repeaters. 

We have grouped our discussion into two subsections: one relating to the quantum memory requirements for a repeater, and one relating to the photonic interface between the quantum memory (stationary) qubits and the photonic (flying) qubits. 

\subsection{Quantum Dots as Quantum Memories}

To evaluate the potential for quantum dot spin qubits to be used as quantum memories in a quantum repeater, one can evaluate them against the first five DiVincenzo criteria. The DiVincenzo criteria are, however, only a rough guide, and to accurately assess whether a technology may be used to produce a working repeater or not, one needs to consider a detailed repeater design, including the specifics of the error correction scheme to be used. Work towards this goal has been done by Jones {\it et al.} \cite{jones_layered_2012} for a quantum computer based on optically controlled quantum dot qubits, but a detailed design for a quantum-dot-based quantum repeater is not yet available. However, from Ref. \cite{jones_layered_2012}, we have a basic idea of the performance required from quantum dot qubits in order to produce a functioning fault-tolerant machine, and at this stage more experimental progress is needed (to provide precise numbers about achievable operation fidelities and times) before a more specific design will be needed to provide a roadmap for further experiments. 

Before we start to consider the details of how a fault-tolerant quantum repeater may be constructed using quantum dots, let us first review how quantum dots may meet the DiVincenzo criteria for quantum memories. 

\subsubsection{DiVincenzo Criterion 1: ``A scalable physical system with well-characterized qubits''}
\label{sec:DVC1}

This criterion imposes two main requirements: that the system being proposed to implement a qubit can be well-described as a quantum two-level system (and therefore that the system has a negligible probability of being found in states besides $\ket{0}$ and $\ket{1}$), and that this system be scalable.

A single quantum dot can trap a single conduction band electron, or a single valence-band (heavy) hole. This can be done deterministically, by embedding a layer of quantum dots in a diode structure -- this is likely the configuration that will be used in a large-scale system. However, many current experiments use stochastic charging of the quantum dots, by placing a layer of $n$-type or $p$-type dopant near the quantum dot layer. 

Regardless of the engineering method used to charge the quantum dots in a sample, the key idea is that a single quantum dot can stably trap a single charge (electron or hole), and the spin state of this charge (which we denote as $\ket{\uparrow}$ and $\ket{\downarrow}$ in the case of an electron\footnote{We use $\ket{\Uparrow}$ and $\ket{\Downarrow}$ to refer to the pseudospin eigenstates of a hole.}) will serve as the qubit, i.e., $\ket{\psi} = \alpha \ket{\uparrow} + \beta \ket{\downarrow}$. We can define the traditional quantum information ``computational basis'' in terms of these eigenstates ($\ket{0} \triangleq \ket{\uparrow}$, and $\ket{1} \triangleq \ket{\downarrow}$), which gives us a single qubit with the notation used in the quantum information literature: $\ket{\psi} = \alpha \ket{0} + \beta \ket{1}$.

In the case of the electron, which is a spin-$\frac{1}{2}$ particle, there are only two spin eigenstates, so an isolated spin seems to easily meet the requirement that the system we choose should have a low probability of being found in a state besides $\ket{\uparrow}$ or $\ket{\downarrow}$. A magnetic field is typically used to split the spin eigenstates in energy. 

A single spin in a quantum dot is, however, not completely isolated: it is part of a larger system (the quantum dot), so there are other eigenstates of the broader system that could potentially be excited. For example, if a photon of an appropriate energy impinges on the quantum dot, it is possible that the photon may be absorbed by the quantum dot, creating an {\it exciton} (an electrostatically-bound electron-hole pair) in the QD. The quantum dot then contains two electrons and a hole (this three-particle set is typically called a {\it trion}), and it is appropriate to then model the system as having transitioned from a state well-described by the spin of just a single electron, to one that consists of two electron spins, and a hole spin. Fortunately in quantum dots, the energy of such an optical transition is large ($\gg k_\text{B} T$, even for room temperature), so the probability for a single quantum dot spin to become a trion without the experimentalist explicitly shining light onto the quantum dot is negligible.  

It is also possible for a charged quantum dot to become uncharged, as the electron (for example) in it tunnels out. One can think of this as a transition to a third state, or just as the loss of the qubit. Fortunately it turns out that quantum dots can stably trap charges for upwards of $20 \ms$ \cite{kroutvar_optically_2004}, so for the timescales of current experiments (which last for at most a few microseconds per run), this is not a major concern. Excitation to, and relaxation from, third (or higher) states is fundamentally connected to decoherence, which we discuss later. 

\subsubsection{DiVincenzo Criterion 2: ``The ability to initialize the state of the qubits to a simple fiducial state''}

In quantum computation, the ability to initialize qubits is crucial for implementing any algorithm, since (in the gate model) algorithms begin by assuming that qubits are in some particular initial state (for example, each qubit being in the state $\ket{0}$). Repeaters have a similar requirement, although depending on the specifics of the physical protocol used to interface the quantum memory with photonic qubits, the initial state might not necessarily be one of the computational basis states ($\ket{0}$ and $\ket{1}$), nor a superposition of them, but some third state. 

For the proposals we discuss in this chapter concerning quantum memories made from spins in optically-active quantum dots, it is sufficient to be able to initialize each qubit in the quantum memory to one of the computational basis states, e.g., $\ket{0}$. 

\begin{figure}[t]
\centering
\includegraphics[width=7cm]{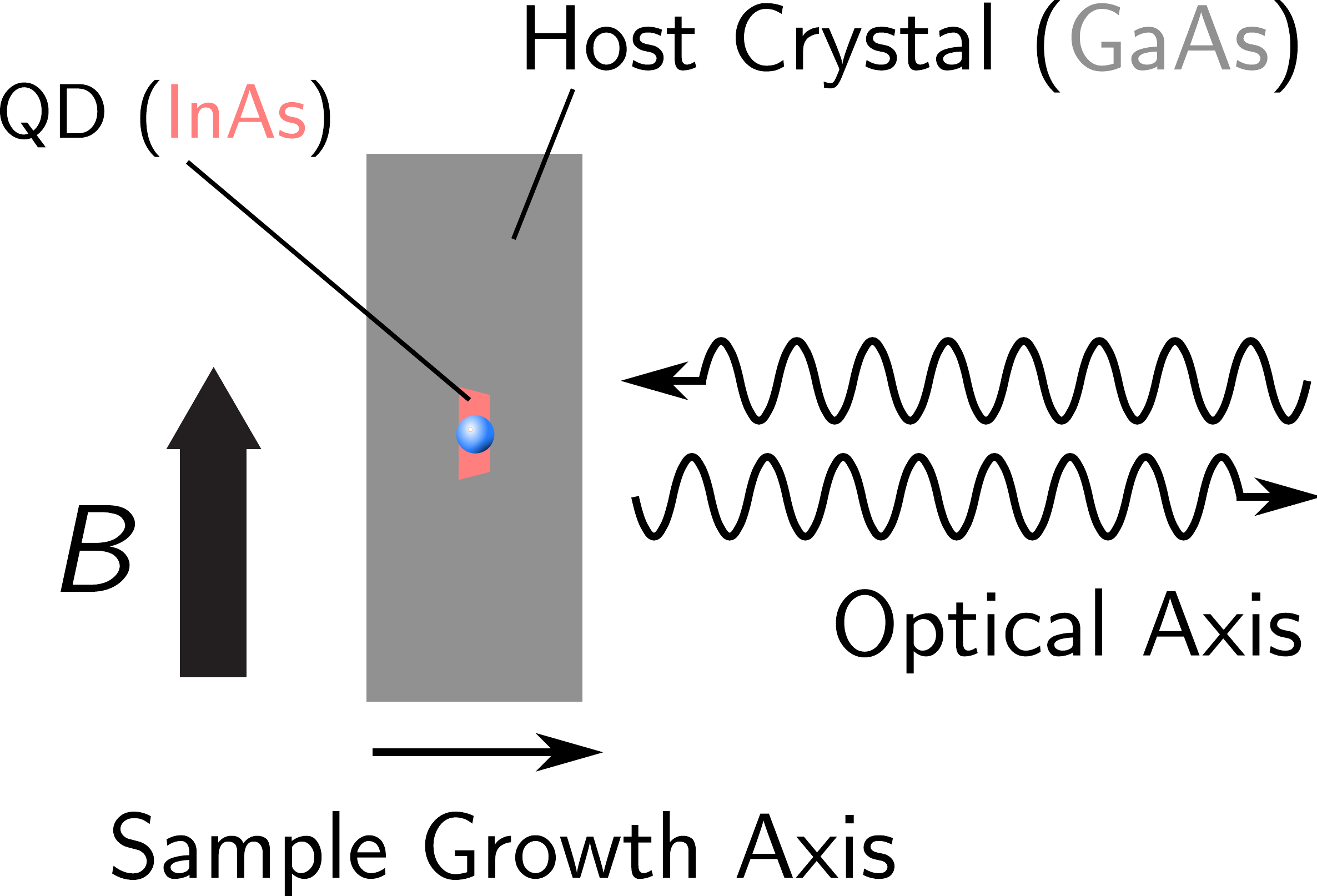}
\caption{Voigt Geometry. The Voigt geometry denotes an experimental setup in which a magnetic field is applied perpendicular to the growth axis of the sample. The optical axis (the axis along which excitation light that impinges on the sample propagates, and the axis along which emitted or reflected light that is collected by a lens propagates) is parallel to the growth axis. Self-assembled InAs quantum dots in a GaAs host crystal are significantly shorter in the growth axis than in either of the in-plane axes. A QD typically has a height (dimension in the growth axis) in the range $\SIrange{1.5}{4}{\nm}$, and a diameter (base length) in the range $\SIrange{20}{40}{\nm}$.}
\label{fig:VoigtGeometry}
\end{figure}

The primary method that is used to perform spin initialization of optically-active quantum dots is optical pumping. This is a technique borrowed from atomic physics \cite{happer_optical_1972}, and was demonstrated for spins in quantum dots in the so-called Voigt geometry by Xu {\it et al.} in 2007 \cite{xu_fast_2007}. The Voigt geometry is the name given to the experimental configuration when the magnetic field is aligned perpendicular to the optical axis and crystal growth axis, as shown in Figure \ref{fig:VoigtGeometry}. This is the geometry in which spin-photon entanglement has been achieved, so it is the geometry we focus on in this review. 

In the Voigt geometry, the first optically excited states of a charged quantum dot are the trion states. Suppose that a quantum dot contains a single electron. If this quantum dot absorbs a photon, it will then contain an electron-hole pair, and the conduction-band electron that was already in the QD, i.e., a trion (as we explained in Section \ref{sec:DVC1}). 

\begin{figure}[t]
\centering
\includegraphics[width=7cm]{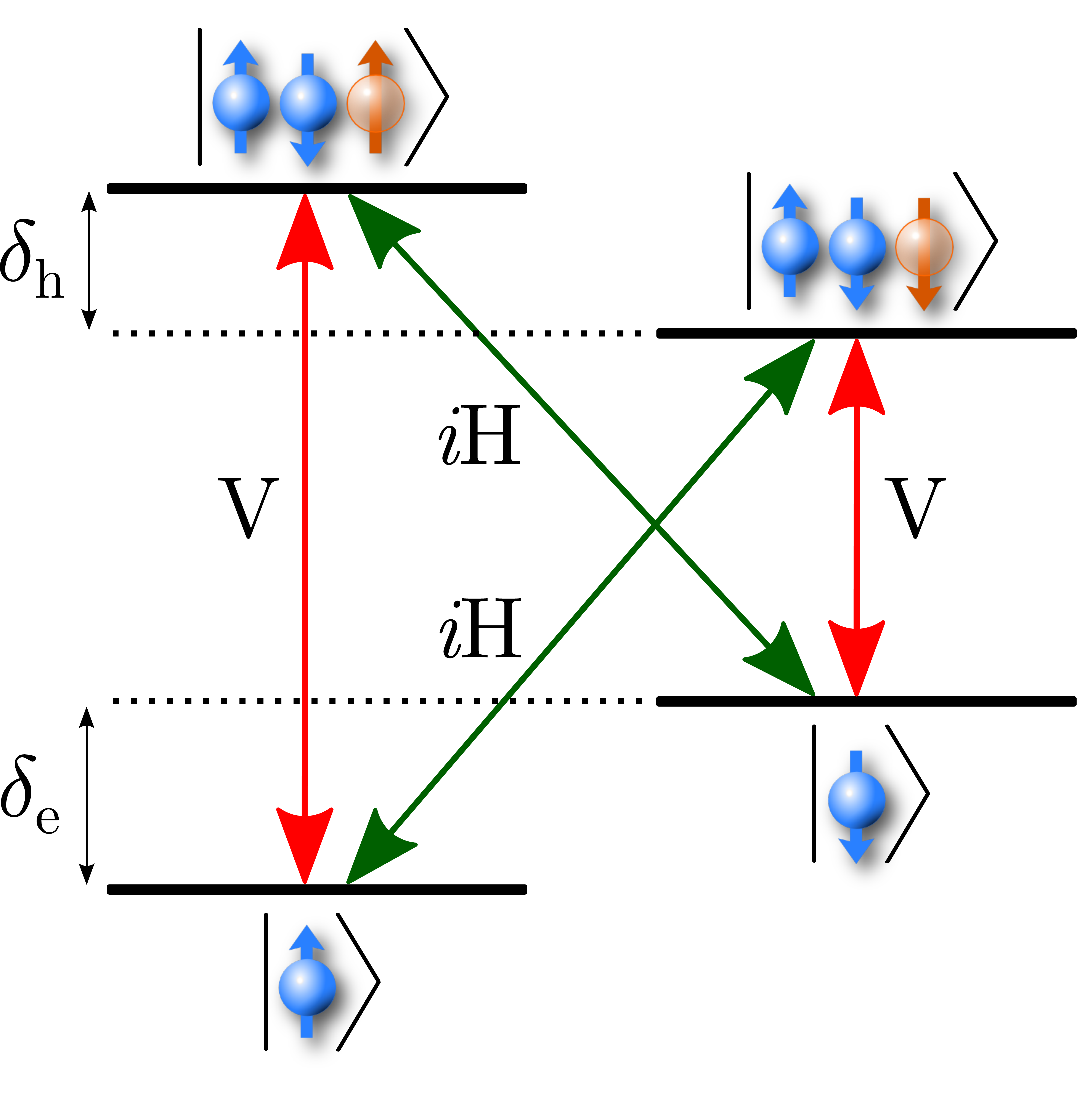}
\caption{Level Diagram and Optical Selection Rules of a Quantum Dot in a Magnetic Field in the Voigt Geometry. When a charged quantum dot is placed in a magnetic field, the electron spin states are split in energy; this Zeeman splitting is denoted in this figure as $\hbar \delta_\text{e}$. The two lowest-energy optical excited states are shown, and also have a Zeeman splitting (due to the interaction of the hole with the magnetic field). Both trion states can decay to either electron spin ground state, with approximately equal probability (oscillator strength). Note that the transitions $\ket{\uparrow} \leftrightarrow \ket{\uparrow\downarrow\Uparrow}$ and $\ket{\downarrow} \leftrightarrow \ket{\uparrow\downarrow\Downarrow}$ have vertical-polarization selection rules, and the transitions $\ket{\uparrow} \leftrightarrow \ket{\uparrow\downarrow\Downarrow}$ and $\ket{\downarrow} \leftrightarrow \ket{\uparrow\downarrow\Uparrow}$ have horizontal-polarization selection rules, but with a $90^\degree$ phase shift.}
\label{fig:QDLevelDiagramVoigt}
\end{figure}

The relevant energy level diagram and optical selection rules for the system in the Voigt geometry are shown in Figure \ref{fig:QDLevelDiagramVoigt}. A feature of this diagram that is relevant to optical pumping, as well as spin rotation and spin-photon entanglement, is that the optically-excited states form two $\mathrm{\Lambda}$ systems with the ground spin states. The fact that the two trion states have allowed optical transitions to both spin ground states is crucial. 

Optical pumping allows the spin to be initialized into one of the two spin ground states on a few nanosecond timescale by applying a narrowband CW laser resonantly on any one of the four optical transitions. 

We will discuss briefly in Section \ref{sec:DVC_measurement} how optical pumping can also be used to perform spin measurement. There are alternatives to spin pumping for initialization, and the one most likely to be used in a large-scale, fault-tolerant system is some form of single-shot, quantum non-demolition (QND) measurement: if one can perform an ideal von Neumann projective measurement on a qubit, then after the measurement the qubit will be in the state $\ket{0}$ or $\ket{1}$, and based on the measurement result one can perform a $\mathtt{NOT}$ gate to flip the spin if needed, and in that way initialize the spin to $\ket{0}$ (or $\ket{1}$, if desired).

\subsubsection{DiVincenzo Criterion 3: ``Long relevant decoherence times, much longer than the gate operation time''}

Spin-based qubits have been considered in many physical systems, since spin is an especially attractive degree of freedom to use for storing quantum information. Not only do spin-$\frac{1}{2}$ particles by definition have only two spin levels (which helps in avoiding the problem of keeping whatever subsystem is being used as a qubit from accidentally exiting into third, fourth, etc., levels), but spin tends to not couple as strongly to uncontrolled degrees of freedom. For example, one could imagine defining a qubit's two states as being two different spatial wavefunctions of an electron. This has the significant disadvantage that not only does one then need to find a way to avoid exiting from the two-state manifold, but also that the wavefunction degree of freedom is significantly affected by Coulomb interactions with nearby charges (i.e., charge noise) \cite{loss_quantum_1998}. The relative insensitivity of the spin degree of freedom to many sources of noise leads to spin qubits having relatively long coherence times, not only in quantum dots, but in other physical systems too. 

In the case of electron spin qubits in self-assembled, optically-active InAs quantum dots formed in GaAs, the $T_2$ coherence time is typically in the range $\SIrange{1}{3}{\us}$; this has been measured for a single quantum dot using a spin echo \cite{hahn_spin_1950} sequence. Hole spin qubits have also been created and their coherence time directly measured using a spin echo sequence; De Greve {\it et al.} measured $T_2 \approx 1.1 \us$ \cite{de_greve_ultrafast_2011} for one such qubit. As the third DiVincenzo criterion says, these $T_2$ values need to be compared to the gate operation times in order to evaluate their suitability.\footnote{Comparing the $T_2$ time to the gate operation times is overly simplistic. In prototype demonstrations of quantum repeaters where the quantum memories are not protected by quantum error correcting codes, then, as we have explained previously, the $T_2$ times need to also be compared to the relevant photon propagation times, with some consideration of entanglement generation heralding success probability. In the case where fault-tolerant quantum-error-corrected memories are to be built, the code and implementation details may call for $T_2$ times much longer than the gate times, but certainly the gate times provide a lower-bound on the requisite $T_2$ times.} 

The existence of optical transitions in the quantum dots is useful for several reasons. The main focus of this book, and of this chapter, is on the interface between photonic qubits and stationary (memory) qubits, and the optical transitions naturally facilitate direct conversion between these two forms of qubits in the quantum dot system. Another advantage has to do with scaling: if we can perform all the operations on our stationary qubits using radiation at optical frequencies, there may be no need for complicated wiring on-chip in order to deliver initialization, control and measurement pulses to specific quantum dots. As Ref. \cite{jones_layered_2012} discusses in some detail, a full quantum processor could potentially be made from a sample that contains no wiring between any of the quantum dots in a large 2D array, where all addressing is performed by beam-steering using micromirrors. The use of optical radiation allows neighbouring qubits to be individually addressed despite being very densely packed; a spacing of $1 \um$ should be sufficient to allow diffraction-limited spots to focus on individual quantum dots with negligible undesired impact on neighbouring qubits. The benefit of optical transitions most relevant to the third DiVincenzo criterion is, however, that gates implemented using optical pulses can be significantly faster than gates implemented using microwave frequency pulses that manipulate the spin ground states directly \cite{de_greve_towards_2013}. 

\subsubsection{DiVincenzo Criterion 4: ``A `universal' set of quantum gates''}

Universal control over a single quantum dot spin qubit has already been demonstrated, both for an electron spin qubit \cite{press_complete_2008}, and for a hole spin qubit \cite{de_greve_ultrafast_2011}. In both cases, a single rotation about the optical axis can be implemented on a timescale of approximately $\SIrange{2}{4}{\ps}$, and a single rotation about the magnetic field (orthogonal) axis is realized by Larmor precession on a timescale of up to $50 \ps$ (depending on the magnitude of the external magnetic field used, and on the spin $g$-factor). A single qubit can be set to an arbitrary position on the Bloch sphere in well under $100 \ps$. The single qubit gate time is thus four orders of magnitude shorter than the $T_2$ coherence time.\footnote{While the single qubit gate time clearly passes the DiVincenzo criterion that it should be much shorter than the $T_2$ time, it is necessary to develop and evaluate a full quantum computer design to be able to properly assess whether the timescales are truly compatible. We focus more on near-term experiments in this chapter, but for a discussion of the requirements in a fault-tolerant quantum computer based on quantum dots, see Ref. \cite{jones_layered_2012}.} In other words, $\sim 10^4$ single qubit operations could be performed on a qubit before it decoheres, provided that a suitable spin echo scheme is used, and under the assumption that the fidelities of the single qubit operations are sufficiently high.\footnote{Currently the fidelities of the single qubit gates limit the number of operations that can be applied to $\ll 100$; in practice, several orders of magnitude improvement in the gate infidelities would be needed to allow a sequence of $10^4$ operations to be usefully applied to a qubit.}\textsuperscript{,}\footnote{Thus far we have avoided mentioning the dephasing time $T_2^*$. However, it is not irrelevant, even when spin echo pulses are used: the single-qubit gate fidelities are closely related to this parameter ($T_2^*$). The dephasing time reflects the (time-averaged) uncertainty about the Larmor precession frequency, and this uncertainty results in errors in single-qubit gates. For example, for rotations (nominally) about the optical axis (induced by picosecond optical pulses), the dephasing processes result in a random, off-axis component on top of the optical-axis rotation, i.e., a random deviation from the ideal behaviour of the gate. This error mechanism can be mitigated if carefully-designed spin-echo-related schemes are used; these methods call for the concatenation of pulses in order to make so-called decoherence-protected gates, but have yet to be realized for quantum dot spin qubits. For the conventional single-qubit gate operations described above, the ratio between the gate operation time and the dephasing time ($T_2^* \approx 1 \ns$ \cite{press_ultrafast_2010}) results in single-qubit gate fidelities that are theoretically limited (by this effect) to $\sim 99.6 \%$ (optical-axis gate) and $\sim 95 \%$ (Larmor gate); these limits are slightly higher than what has been measured experimentally \cite{de_greve_ultrafast_2011}.} 

For all experiments that have been performed so far, and all those likely to be performed in the near future, the time required to perform single qubit operations does not considerably affect the fidelity of the output state, so long as a spin echo refocussing pulse is used. The dephasing time $T_2^*$, which is the relevant decoherence timescale when a spin echo pulse is not used, is approximately $1~\ns$ for electron spins \cite{press_ultrafast_2010}. The $T_2^*$ time is thus only roughly an order of magnitude larger than the single-qubit gate time. 

Although the $T_2$ time is sufficiently long that the finiteness of the time taken to perform single qubit gates is generally not a dominant cause of error (infidelity), the $T_2$ time is nevertheless an important experimental parameter in current experiments exploring spin-photon and spin-spin entanglement with quantum dots. As we have mentioned earlier in this chapter, in even the simplest entanglement distribution experiments, the coherence time of the memory needs to be long compared to the time taken for photons to propagate. For example, if one intends to entangle two spins in remote quantum dots, the two cryostats should be connected by a fibre length that is substantially less than $L_\text{max} = T_2 \frac{c}{2 \cdot n_\text{core}}$, which for $T_2 \approx 3~\us$, yields $L_\text{max} \approx 555~\meter$. This is a perfectly reasonable value for the purposes of laboratory proof-of-principle demonstrations, but clearly an extension to the intrinsic coherence time, or the development of an error-protected quantum memory, will be necessary to perform long-distance experiments. 

Single qubit gates alone are not universal for computation, so the second part of this DiVincenzo criterion calls for the demonstration of a scalable two-qubit (entangling) gate, for example, a $\mathtt{CNOT}$ gate. There are several proposals for how to implement such a gate for quantum dot spin qubits \cite{imamoglu_quantum_1999,piermarocchi_optical_2002,quinteiro_long-range_2006,ladd_simple_2011,puri_two-qubit_2012}, but there have been no experimental demonstrations thus far. Kim {\it et al.} \cite{kim_ultrafast_2011} showed that one can perform a two-qubit gate that is mediated by an always-on exchange interaction between two adjacent quantum dots in a quantum dot molecule structure, but unfortunately this approach is not scalable beyond a few qubits. One of the major outstanding experimental challenges for optically-active quantum dot spin qubits is the demonstration of a scalable, fast, high-fidelity two-qubit gate, which is a prerequisite for the implementation of error correction codes.

\subsubsection{DiVincenzo Criterion 5: ``A qubit-specific measurement capability''}
\label{sec:DVC_measurement}

As a method for qubit initialization, optical pumping performs well. However, this method is also used to perform qubit readout in most\footnote{For example, the recent demonstrations of spin-photon entanglement from three different groups all used this method \cite{de_greve_quantum-dot_2012,de_greve_complete_2013,gao_observation_2012,schaibley_demonstration_2013}.} optical quantum dot spin qubit experiments \cite{de_greve_ultrafast_2013}. The basic principle of this type of readout is that during optical pumping, the quantum dot will emit a single photon on the branch of the $\mathrm{\Lambda}$ system that is not being pumped (e.g., $\ket{\uparrow\downarrow\Downarrow} \rightarrow \ket{\uparrow}$) if and only if the spin was in one particular state ($\ket{\downarrow}$), but the quantum dot will emit no photons along that branch if the spin was in the other state ($\ket{\uparrow}$). There are two major disadvantages to this optical pumping procedure regarding its use for readout. The most important disadvantage, from the perspective of current experiments, is that per experimental run\footnote{A single run may be a sequence of events such as: 1.) Initialize the spin, 2.) Perform one or more rotation gates on the spin, 3.) Measure the spin.}, at most a single photon will be emitted indicating the spin is in a particular state. Since the overall collection and detection efficiency is small (typically less than $0.1\%$), it is necessary to re-run a particular experiment many times in order to obtain a reasonable signal-to-noise ratio. In the sense that it is necessary to repeat the experiment multiple times to obtain an average measurement outcome, this type of readout does not implement a ``single-shot'' measurement, and, for example, cannot be used to detect quantum jumps (or other phenomena associated with single quantum trajectories). 

The second disadvantage of the spin readout based on optical pumping fluorescence is that regardless of the measurement outcome, the qubit ends up in one particular state (for example, $\ket{\uparrow}$). In this sense the method does not perform a ``quantum non-demolition'' (QND) measurement, which we use here to mean just that the measurement does not act as a textbook von Neumann projective measurement. 

There are proposals for implementing scalable single-shot QND measurements, in both the Voigt and Faraday geometries. In the Faraday geometry, the existence of a cycling transition allows a fluorescence-based measurement \cite{delteil_observation_2014} that is impossible in the Voigt geometry, but unfortunately the single-qubit gate mechanism used in the work we have described in the previous subsection relies on the selection rules in the Voigt geometry. As yet there have been no demonstrations of single-qubit gates and single-shot QND readout in the same experiment. The spin-dependent Faraday- or Kerr-rotation of a probe pulse, which has been demonstrated in multi-shot experiments \cite{atature_observation_2007,berezovsky_nondestructive_2006}, may plausibly lead to a single-shot readout in the Voigt geometry. In the Faraday geometry, besides the cycling transition, one may also use the spin-dependent Faraday or Kerr rotations, or a polariton-based mechanism \cite{puri_single-shot_2014}. Single-shot readout using a cycling transition in a quantum dot molecule has been demonstrated in the Faraday geometry \cite{vamivakas_observation_2010}, but is yet to be realized in the Voigt geometry. 

\subsection{Quantum Dots as Photon Sources}

The suitability of optically-controlled quantum dot spins as quantum memories can be evaluated against the first five DiVincenzo criteria. To evaluate their use as building blocks for a quantum repeater, we need to consider the final two DiVincenzo criteria. We will first consider DiVincenzo Criterion 7: ``The ability to faithfully transmit flying qubits between specified locations''. One can imagine using electrons, or some other matter particles, as flying qubits, but this seems exceptionally difficult for even moderate macroscopic distances (i.e., on the order of meters). Therefore nearly all proposals for flying qubits consider optical-frequency photons, either in free-space or in optical fibre: these photons can encode quantum information in degrees of freedom that are very robust against decoherence, and they can be transmitted over relatively long distances with relatively low loss. 

The use of quantum dots as photon sources doesn't directly address either DiVincenzo Criteria 6 or 7, but is related to both, and is an important area of research in the quantum dot community, both for its relevance to quantum repeaters, and other aspects of quantum-optics-based quantum information technology. 

Quantum dots have been shown to be outstanding single-photon sources, i.e., they can produce single photons on demand (with either electrical or optical triggering). Considerable effort has been expended over the past 15 years in making quantum-dot-based single-photon sources that have very low $g^{(2)}(0)$ values, and good indistinguishability. Both of these are important parameters for quantum repeaters. It is easy to see why a non-zero $g^{(2)}(0)$ value negatively affects the entanglement generation protocol we have described: if either of two quantum dots that are to be entangled have non-zero $g^{(2)}(0)$, then it is possible that the detectors measure a double-click event (which should herald entanglement between the two quantum dots) even though no photon arrived at the detectors from one of the quantum dots. Therefore some of the heralded events will not actually correspond to cases where the quantum dots are in the target entangled state, and this will result in an overall reduction in the fidelity of the entangled state. Imperfect indistinguishability of photons also results in a reduction in state fidelity, and in reduced efficiency of entanglement generation.

Single-photon sources are sought after not only for quantum repeaters (as part of a spin-photon interface), but in their own right for use in quantum key distribution ({\it sans} quantum repeaters) and linear-optics-based quantum computing \cite{kok_linear_2007}. In BB84-based QKD with single-photon sources, it is desirable to have $g^{(2)}(0)$ be as low as possible, since if the source emits more than one photon per time slot, it may be possible for an eavesdropper to gain information without being detected.\footnote{Decoy-state methods to allow the use of attenuated coherent light sources, rather than single-photon sources, in BB84-based QKD applications have been remarkably successful \cite{lo_secure_2014}. The rise in these methods has reduced the desire for single-photon sources for BB84 implementations. Good single-photon sources are however still highly desirable for linear-optics-based quantum computing \cite{jennewein_single-photon_2011,varnava_how_2008}.} In linear optical quantum computing, it is important for the photonic qubits to interfere with one another, so indistinguishability is crucial. 

The inhomogeneity of quantum dots (different quantum dots tend to have different optical emission wavelengths, and different linewidths) is a major drawback; photon indistinguishability is a prerequisite for interference, and spin-spin entanglement protocols such as Simon-Irvine rely centrally on Hong-Ou-Mandel-style interference. There have been demonstrations of interference between photons emitted from different, remote quantum dots \cite{patel_two-photon_2010,gao_quantum_2013,gold_two-photon_2014}, but certainly for large-scale use, the lack of homogeneity is an outstanding problem. One main approach to solving this problem is to tackle it directly through improvements in sample growth and fabrication; however, it may also be possible to use frequency conversion \cite{zaske_visible--telecom_2012,pelc_downconversion_2012,de_greve_quantum-dot_2012} to help achieve interference of photons emitted from quantum dots at disparate wavelengths \cite{ates_two-photon_2012}. 

There has also been a large research effort in developing entangled-photon-pair sources using quantum dots.\footnote{By this we mean devices that emit (preferably on-demand, using either an optical or an electrical trigger) two photons at a time, and these photons are entangled with each other. For example, a common type of entangled-photon-pair source is one that emits polarization-entangled photons, i.e., it generates two-photon quantum states such as $\ket{\psi} = \frac{1}{\sqrt{2}} \left( \ket{\text{H}}_\text{A} \ket{\text{V}}_\text{B} + \ket{\text{V}}_\text{A} \ket{\text{H}}_\text{B}  \right)$.} Since a proposal in 2000 \cite{benson_regulated_2000}, there have been multiple demonstrations \cite{akopian_entangled_2006,dousse_ultrabright_2010,juska_towards_2013}. The main disadvantage of these entangled-pair sources compared to spontaneous parametric downconversion in nonlinear crystals \cite{pan_multiphoton_2012} is the need for the quantum dot samples to be kept at cryogenic temperatures during operation. Sufficiently reducing the fine-structure splitting in quantum dots (which otherwise leaks which-path information, resulting in reduced fidelity of the target entangled two-qubit state) is non-trivial, and needs careful attention \cite{gilchrist_quantum_2007}. In any case, besides the uses of entangled-photon sources to demonstrate entanglement-based QKD protocols (such as Ekert91), and small-scale linear optical quantum computing, there is an important potential use for such sources in quantum repeater networks: the proposal of Jones {\it et al.} \cite{jones_high-speed_2013} calls for the use an entangled-photon source between each pair of quantum repeater nodes. If frequency conversion is not used in an initial demonstration of this Jones {\it et al.} protocol, then wavelength matching could perhaps be performed more easily between the repeater nodes and the entangled-photon-pair source if they are both made using quantum dots. If frequency conversion to the telecom wavelength is performed, then a telecom-wavelength entangled-photon-pair source would be needed; besides work on growth of quantum dots that natively emit at telecom wavelengths, there are also other sources of entangled photons at telecom wavelengths that are available; for a recent example, see Ref. \cite{jin_pulsed_2014}. 

One final aspect of quantum dot single-photon sources that is relevant to the spin-photon interface are demonstrations of pulsed resonant coherent excitations of optical transitions in quantum dots. Zrenner {\it et al.} \cite{zrenner_coherent_2002} showed Rabi oscillations between the crystal ground state of a neutral quantum dot, and an exciton state, and Pelc {\it et al.} \cite{pelc_downconversion_2012} showed Rabi oscillations between a spin ground state of a charged quantum dot, and one of the trion states, including downconversion of the measured fluorescence to the telecom wavelength. These results show that it is possible to optically excite the exciton or trion states with high probability, which is useful for the start of a protocol to generate spin-photon entanglement. 

\subsection{Entanglement Between a Spin in a Quantum Dot and an Emitted Photon}

In the previous subsection we have summarized how research on quantum dots as single photon and entangled photon pair sources may bear on the use of quantum dots in quantum repeater networks. A physical system that can act as a good single-photon source has some promising attributes that may also allow it to perform as a good spin-photon interface, but we have not yet described the other necessary conditions. 

The final DiVincenzo criterion for us to consider is number 6: ``The ability to interconvert stationary and flying qubits''. As we have described earlier, one popular technique to generate entanglement between stationary qubits mediated by flying qubits is to use the Simon-Irvine protocol, or a variant thereof. This protocol requires the generation of pairs of stationary and flying qubits that are entangled. The sixth DiVincenzo criterion suggests a requirement more along the lines of converting a stationary qubit into a flying qubit, and then converting that flying qubit into a stationary qubit at a different location, but the literal interpretation of this as necessarily being a direct physical process is overly restrictive: so long as you can distribute entanglement over long distances using flying qubits, you can transfer quantum information using quantum teleportation. 

\begin{figure}[t]
\centering
\includegraphics[width=5cm]{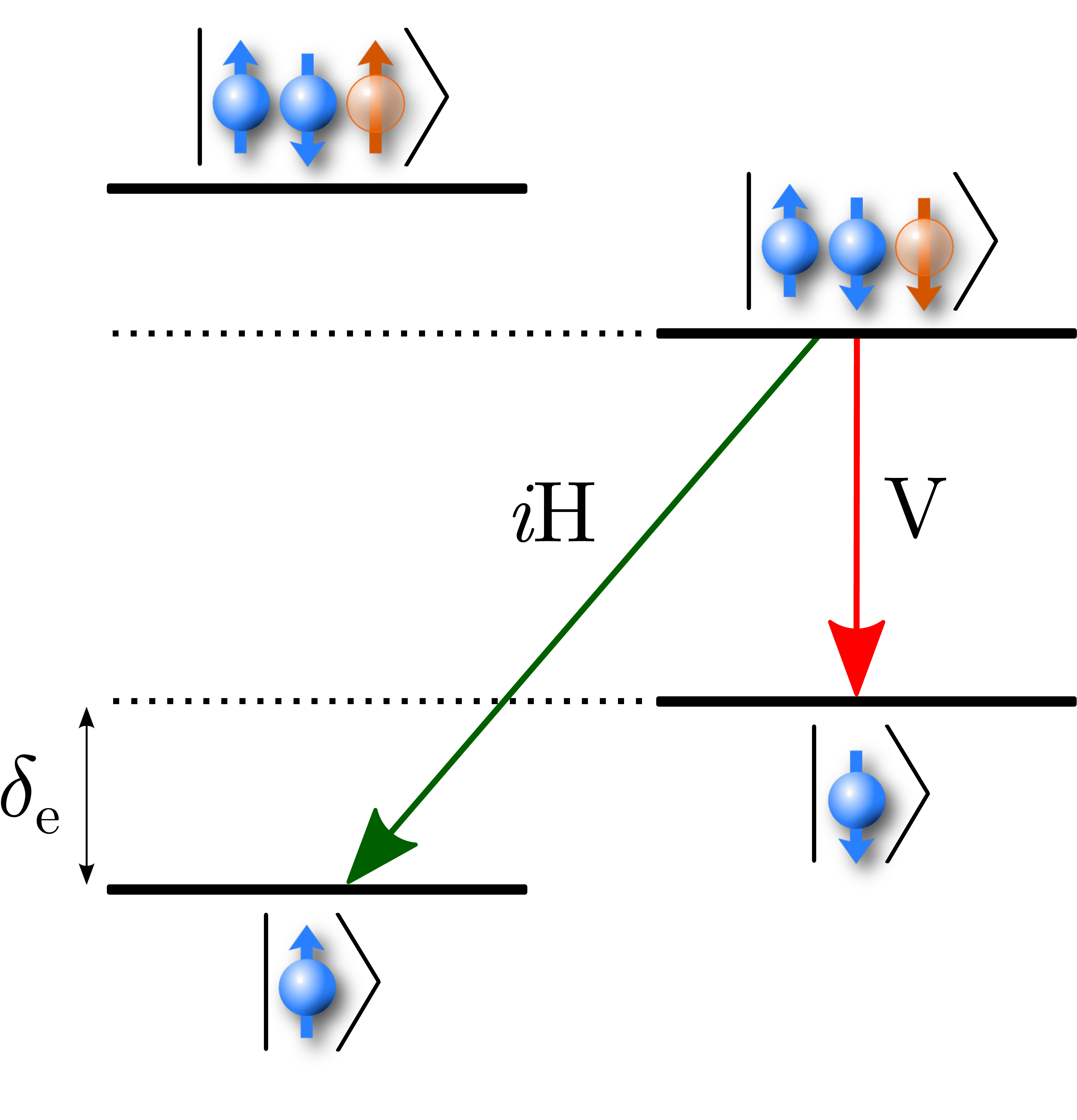}
\caption{Spin-Photon Entanglement Generation from a Charged Quantum Dot in a Voigt-geometry External Magnetic Field. When the trion state $\ket{\uparrow\downarrow\Downarrow}$ decays via spontaneous emission, it does so with with an equal amplitude of decaying to the state $\ket{\uparrow}$ or the state $\ket{\downarrow}$. However, the polarization selection rules for the $\ket{\uparrow\downarrow\Downarrow} \rightarrow \ket{\uparrow}$ and the $\ket{\uparrow\downarrow\Downarrow} \rightarrow \ket{\downarrow}$ decays are different. Furthermore, the $\ket{\uparrow}$ and $\ket{\downarrow}$ states are not energy-degenerate: they are split by a Zeeman energy $\hbar \delta_\text{e}$. Therefore the decay of the trion state results in the generation of an entangled state, where there is hyperentanglement between the electron ground spin states and two properties of the emitted single photon: its energy, and its polarization. The state produced can be written as $\ket{\psi} = \frac{1}{\sqrt{2}} \left( \ket{\uparrow} \ket{i \text{H}, \hbar ( \omega + \delta_\text{e} ) } + \ket{\downarrow} \ket{\text{V}, \hbar \omega } \right)$, where $\hbar \omega$ is the energy of the $\ket{\uparrow\downarrow\Downarrow} \leftrightarrow \ket{\downarrow}$ optical transition.}
\label{fig:QDLevelDiagramVoigtSpinPhoton}
\end{figure}

The Simon-Irvine protocol is an elegant way to distribute entanglement, and is very well-suited to quantum dots, since charged quantum dots provide a direct mechanism for generating entanglement between a stationary qubit and a flying qubit \cite{economou_unified_2005}. Consider the energy level diagram describing the relevant spin ground states, the first optically-excited states (trions), and the relevant optical selection rules for a charged quantum dot, in a Voigt-geometry magnetic field. Figure \ref{fig:QDLevelDiagramVoigtSpinPhoton} shows the four-level diagram, and the optical selection rules for the allowed transitions from the trion state $\ket{\uparrow\downarrow\Downarrow}$. We denote as $\hbar \omega$ the energy of the $\ket{\uparrow\downarrow\Downarrow} \leftrightarrow \ket{\downarrow}$ optical transition. If the system begins in the trion state $\ket{\uparrow\downarrow\Downarrow}$, then once this state decays (which takes on average approximately $1~\ns$ if the quantum dot emission is not enhanced by an optical cavity), the following spin-photon entangled state is produced:

\begin{equation}
  \ket{\psi} = \frac{1}{\sqrt{2}} \left( \ket{\uparrow} \ket{i \text{H}, \hbar ( \omega + \delta_\text{e} ) } + \ket{\downarrow} \ket{\text{V}, \hbar \omega } \right).
\label{eq:SpinPhotonEntanglement}
\end{equation}

This state is hyperentangled, in the sense that the spin qubit is entangled with two different properties of the emitted photon: both its polarization and its energy. Entanglement between the spin and the photon polarization, and between the spin and the photon energy, have both been experimentally verified.

\subsubsection{General Standards of Experimental Proof of Entanglement}

The state in Eq. \ref{eq:SpinPhotonEntanglement} is entangled, since it is a non-separable state \cite{nielsen_quantum_2011}. However, in an experiment, we would like to be able to prove that a process we claim produces an entangled state actually does so. 

Two commonly-used methods for experimentally proving entanglement of a two-qubit state are:

\begin{itemize}
  \item Option 1: Perform, in two orthogonal bases, measurements that yield conditional probabilities, and show that ``strong'' correlations exist in both bases. These results can typically provide a lower bound on the fidelity of the state that is produced.
	\item Option 2: Perform full quantum state tomography to reconstruct the density matrix of the two-qubit state that is produced. This allows a direct calculation of the fidelity of the state that is produced.
\end{itemize}

The measurements used in the first option are generally a subset of those required in the second option; since full state tomography yields the state fidelity $F$ rather than just a bound on it, full tomography is preferable. Not only that, but knowledge of the full reconstructed density matrix may aid in debugging experimental imperfections or refining theoretical predictions of the output state. 

The fidelity $F$ is defined, in the same way we have used it earlier in this chapter, as the overlap between the ``ideal'' (or ``target'') state (for example, $\ket{\psi}$ in Eq. \ref{eq:SpinPhotonEntanglement}) and the measured density matrix $\rho$ of the state that was actually produced in an experiment. Formally, $F \triangleq \left\langle \psi |\rho \middle| \psi \right\rangle$, and $F \in [0,1]$. $F=1$ reflects that the density matrix is exactly that of the pure state $\ket{\psi}$, i.e., $\rho = \ket{\psi}\bra{\psi}$. The measured density matrix, and hence the measured fidelity, will reflect both deviations of the two-qubit state that is produced from the ideal two-qubit state, and errors in the measurement of the two-qubit state. Measurement errors thus place a bound on the maximum observable fidelity of a produced two-qubit state. 

Using Option 1, several (but not all) elements of the density matrix $\rho$ will be obtained, and this will allow the estimate of a lower bound on $F$. For both Option 1 and Option 2, an estimate of $F>0.5$ indicates that the produced state is entangled, assuming that the ideal state $\ket{\psi}$ is maximally entangled. A classical density matrix (one with no off-diagonal elements) can yield a fidelity that is at most $0.5$. 

Related to this threshold for observing entanglement is the notion of a threshold for entanglement purification: entanglement purification can only yield an output state with higher fidelity than the fidelities of the input states if the input state fidelity is greater than $0.5$ \cite{bennett_purification_1996}. In the quantum repeater design of D\"ur {\it et al.} \cite{dur_quantum_1999}, which uses entanglement purification, the threshold is in fact even higher, to compensate for imperfect operations. In general, higher intrinsic fidelities of spin-photon and spin-spin entangled state generation will lead to overall reductions in the resources required to produce distributed, high-fidelity entangled states, which is the goal of a quantum repeater network. 

For Option 1, the methods sections of References \cite{blinov_observation_2004,togan_quantum_2010,de_greve_quantum-dot_2012} explain how to obtain the fidelity bound from several conditional probability measurements. The details of how to perform two-qubit density matrix reconstruction (using the example of spin-photon entanglement) is covered in detail in the supplementary information to Ref. \cite{de_greve_complete_2013}. 

The observation of ``strong'' correlations between measurement results in multiple orthogonal bases is an experimental signature of entanglement. Here we give some simple pedagogical examples to explain what we mean by this, which should aid in gaining intuition about experimental evidence of entanglement. 

Suppose we have two qubits, labeled A and B. Let's suppose both qubits are spin qubits. We can write an entangled pair of such qubits as follows:\footnote{Note that EPR (Bell) states, such as $\ket{\phi^\text{EPR}}_\text{AB}$, are maximally entangled. Moreover, entanglement is monogamous, i.e., if two qubits (A and B) are maximally entangled, it is impossible for a third qubit (C) to become entangled with A or B without reducing the amount of entanglement between A and B \cite{horodecki_quantum_2009}. This lies at the heart of the security of the Ekert91-based QKD protocols.}

\begin{equation}
  \ket{\phi^\text{EPR}}_\text{AB} = \frac{1}{\sqrt{2}} \ket{\uparrow\uparrow}_\text{AB} + \frac{1}{\sqrt{2}} \ket{\downarrow\downarrow}_\text{AB}.
\end{equation}

The general description of a two-qubit state, which may be a mixture of pure states, is given by a $4 \times 4$ density matrix. For example, the density matrix representing $\ket{\phi^\text{EPR}}$ is:

\begin{align}
  \rho_{\ket{\phi^\text{EPR}}} &= \ket{\phi^\text{EPR}} \bra{\phi^\text{EPR}} \\
	                             &= \frac{1}{2}
															    \begin{pmatrix}
															      1 & 0 & 0 & 1 \\
																		0 & 0 & 0 & 0 \\
																		0 & 0 & 0 & 0 \\
																		1 & 0 & 0 & 1
															    \end{pmatrix}.
\end{align}

Now suppose that you have a system that is in a mixed state, represented by a density matrix $\rho_\text{m}$, that is a mixture of unentangled two-qubit states. If you perform measurements on the qubits A and B of this system in just one basis, it is possible that your mixed system may yield the same correlations between measurement results for A and B as those that you would obtain by measuring the the entangled state $\rho_{\ket{\phi^\text{EPR}}}$. For example, suppose you perform measurements in the basis $\left\{ \ket{\uparrow} , \ket{\downarrow} \right\}$. If we have the state $\ket{\phi^\text{EPR}}$, and measure qubit A and get the result $\uparrow$, then when we measure qubit B, we will get the result $\uparrow$ with 100\% probability, i.e., $\text{Pr}\left[B=\uparrow \middle| A=\uparrow \right] = 1$. It is easy to show that if the result of the measurement of qubit A is $\downarrow$, then the result of the measurement of qubit B will be $\downarrow$ with 100\% probability, i.e., $\text{Pr}\left[B=\downarrow \middle| A=\downarrow \right] = 1$. In this sense the measurement results are perfectly correlated. 

Now imagine we have the following classical probabilistic state:

\begin{align}
  \rho_\text{m} &= \frac{1}{2} \ket{\uparrow\uparrow} \bra{\uparrow\uparrow} + \frac{1}{2} \ket{\downarrow\downarrow} \bra{\downarrow\downarrow} \label{eq:EntanglementProof_rhom} \\
	              &= \frac{1}{2}
								   \begin{pmatrix}
								     1 & 0 & 0 & 0 \\
									   0 & 0 & 0 & 0 \\
									   0 & 0 & 0 & 0 \\
										 0 & 0 & 0 & 1
                   \end{pmatrix}.
\end{align}

If we perform the same measurements on preparations of this state, we will obtain exactly the same correlations: $\text{Pr}\left[B=\uparrow \middle| A=\uparrow \right] = 1$ and $\text{Pr}\left[B=\downarrow \middle| A=\downarrow \right] = 1$. 

Now imagine that we perform measurements in the orthogonal basis\footnote{Here we use the definitions $\ket{\rightarrow} \triangleq \frac{1}{\sqrt{2}} \left( \ket{\uparrow} + \ket{\downarrow} \right)$ and $\ket{\leftarrow} \triangleq \frac{1}{\sqrt{2}} \left( \ket{\uparrow} - \ket{\downarrow} \right)$.} $\left\{ \ket{\rightarrow} , \ket{\leftarrow} \right\}$. We can rewrite the state $\ket{\phi^\text{EPR}}$ in this basis, to make it easy to intuitively see what measurement correlations we will obtain. In the $\left\{ \ket{\rightarrow} , \ket{\leftarrow} \right\}$ basis, $\ket{\phi^\text{EPR}} = \frac{1}{\sqrt{2}} \left( \ket{\rightarrow}_\text{A} \ket{\rightarrow}_\text{B} + \ket{\leftarrow}_\text{A} \ket{\leftarrow}_\text{B} \right)$. You can calculate, or using this convenient form of the state, just note that the correlations of $\ket{\phi^\text{EPR}}$ when measured in the $\left\{ \ket{\rightarrow} , \ket{\leftarrow} \right\}$ basis will be: $\text{Pr}\left[B=\rightarrow \middle| A=\rightarrow \right] = 1$ and $\text{Pr}\left[B=\leftarrow \middle| A=\leftarrow \right] = 1$. 

However, if we perform $\left\{ \ket{\rightarrow} , \ket{\leftarrow} \right\}$ basis measurements on the state $\rho_\text{m}$ as given in Eq. \ref{eq:EntanglementProof_rhom}, we will not observe these same correlations. Instead, we will see no correlations between measurement outcomes on systems A and B:

\begin{align}
  \text{Pr}\left[B=\rightarrow \middle| A=\rightarrow \right] &= \frac{1}{2} \\
	\text{Pr}\left[B=\leftarrow \middle| A=\rightarrow \right] &= \frac{1}{2} \\
	\text{Pr}\left[B=\rightarrow \middle| A=\leftarrow \right] &= \frac{1}{2} \\
	\text{Pr}\left[B=\leftarrow \middle| A=\leftarrow \right] &= \frac{1}{2}.
\end{align}

This gives us some intuition for why it is necessary to perform measurements in multiple orthogonal bases in order to verify that a state is entangled: if one only performs measurements in a single basis, the measurement correlations can be the same from a classical probabilistic state as those from an entangled state. However, if a state yields strong correlations in measurements performed in two orthogonal bases, this is considered experimental proof that the state being measured is entangled. The computation of fidelities of reconstructions of the density matrix allows us to formalize this intuition, and be able to use a quantitative separator to distinguish entangled from unentangled states. 

\subsubsection{Demonstrations of Spin-Photon Entanglement with Quantum Dots}

We have provided some intuition for how a single charged quantum dot in a Voigt-geometry magnetic field can be used to generate a two-qubit entangled state, consisting of spin (stationary qubit) and a photon (flying qubit), in the state $\ket{\psi} = \frac{1}{\sqrt{2}} \left( \ket{\uparrow} \ket{i \text{H}, \hbar ( \omega + \delta_\text{e} ) } + \ket{\downarrow} \ket{\text{V}, \hbar \omega } \right)$. We have also briefly outlined current standards for evaluating whether results from an experiment support entanglement having been generated or not. 

Thus far three groups have provided evidence of spin-photon entanglement generation using charged quantum dots. The first two experiments, published jointly in 2012, showed evidence for entanglement between spin and photon polarization, and between spin and photon energy respectively \cite{de_greve_quantum-dot_2012,gao_observation_2012}. A report from Schaibley {\it et al.} \cite{schaibley_demonstration_2013} also showed evidence of entanglement between a quantum dot spin, and photon polarization. All three of these reports produced bounds on the state fidelity by calculating the conditional probabilities for measurements in two orthogonal bases. In a follow-up \cite{de_greve_complete_2013} to their first paper \cite{de_greve_quantum-dot_2012}, De Greve {\it et al.} showed results from a full tomographic reconstruction of the density matrix, yielding strong experimental proof that the entangled state produced by a charged quantum dot is $\ket{\psi} = \frac{1}{\sqrt{2}} \left( \ket{\uparrow} \ket{i \text{H}, \hbar ( \omega + \delta_\text{e} ) } + \ket{\downarrow} \ket{\text{V}, \hbar \omega } \right)$. All the experiments we have mentioned so far in this section (Refs. \cite{de_greve_quantum-dot_2012,gao_observation_2012,schaibley_demonstration_2013,de_greve_complete_2013}) work in quite similar ways. We will focus in particular on the experiments by De Greve {\it et al.} \cite{de_greve_quantum-dot_2012,de_greve_complete_2013}, but the basic concept of how the entanglement generation and verification is performed shares many common aspects with the other works. 

The high-level procedure that is carried out is as follows:

\begin{enumerate}
  \item The quantum dot is prepared in the state $\ket{\downarrow}$ by a combination of optical pumping and, depending on the particular experiment, a $\pi$ rotation operation (that flips the spin from $\ket{\uparrow}$ to $\ket{\downarrow}$). \\
	\item A pulse that drives the $\ket{\downarrow} \leftrightarrow \ket{\uparrow\downarrow\Downarrow}$ transition is applied, with the goal of setting the quantum dot to be in the state $\ket{\uparrow\downarrow\Downarrow}$. \\
	\item The state $\ket{\uparrow\downarrow\Downarrow}$ spontaneously decays and emits a photon, which results in the creation of the spin-photon entangled state $\ket{\psi} = \frac{1}{\sqrt{2}} \left( \ket{\uparrow} \ket{i \text{H}, \hbar ( \omega + \delta_\text{e} ) } + \ket{\downarrow} \ket{\text{V}, \hbar \omega } \right)$. \\
	\item Now that the entangled state has been produced, we seek to measure it. First we measure the state of the photon. In the case of polarization, this is done by using a polarizer and a single-photon detector: if a photon is detected, the photon must have been of the polarization that the polarizer transmits. \\
	\item Next we measure the state of the spin. This is done by performing an optional spin rotation (depending on which basis we want to measure the spin in), and then optical pumping again. A different single-photon detector is used to record if a photon is emitted. The detection or non-detection of a photon by this detector provides the spin measurement result. \\
\end{enumerate}

This describes just a single run of an experiment; for a single choice of measurement bases for the spin and the photon, this is repeated many times. The correlation between photon detections at the two different detectors during the same run of the experiment allows us to determine the conditional probability between a photon polarization measurement outcome and a spin measurement outcome. This whole procedure is then repeated for several different measurement bases, so that at least eight conditional probabilities for different orthogonal measurement outcomes can be determined. 

\begin{figure}[t]
\centering
\includegraphics[width=12cm]{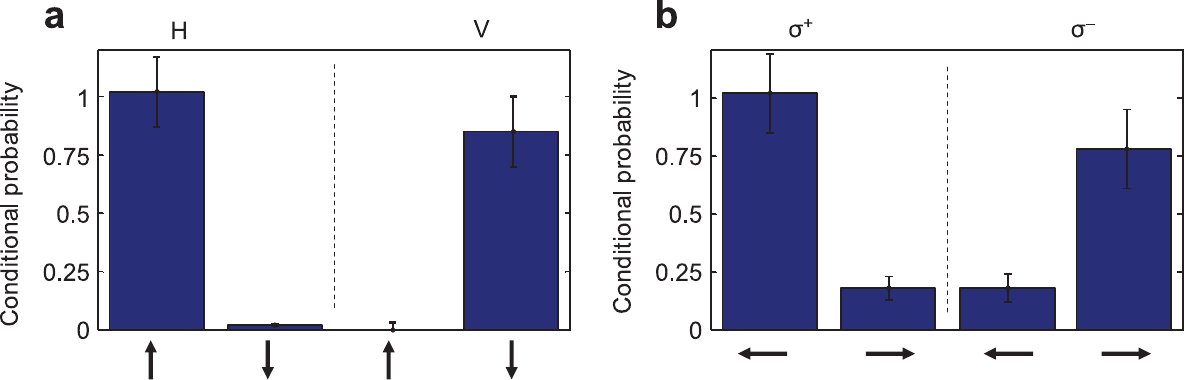}
\caption{Derived and reprinted with permission from De Greve {\it et al.} \cite{de_greve_quantum-dot_2012}. The left panel (a) shows the conditional probabilities when measurements were performed in the $\left\{ \ket{\uparrow} , \ket{\downarrow} \right\}$ basis for the spin, and in the $\left\{ \ket{\text{H}} , \ket{\text{V}} \right\}$ basis for the photon. The right panel (b) shows the conditional probabilities when measurements were performed in the $\left\{ \ket{\rightarrow} , \ket{\leftarrow} \right\}$ basis for the spin, and in the $\left\{ \ket{\sigma^+} , \ket{\sigma^-} \right\}$ basis for the photon.}
\label{fig:nat_correlations}
\end{figure}

Figure \ref{fig:nat_correlations} shows the conditional probabilities obtained by De Greve {\it et al.} \cite{de_greve_quantum-dot_2012}. These conditional probabilities are given as the probability of a spin measurement outcome, given a photon polarization measurement outcome. For example, from the first panel, we can read that:

\begin{align}
  \text{Pr}\left[ \text{spin}=\uparrow \middle| \text{photon}=\text{H} \right] &\approx 1 \\
	\text{Pr}\left[ \text{spin}=\downarrow \middle| \text{photon}=\text{H} \right] &\approx 0 \\
	\text{Pr}\left[ \text{spin}=\uparrow \middle| \text{photon}=\text{V} \right] &\approx 0 \\
	\text{Pr}\left[ \text{spin}=\downarrow \middle| \text{photon}=\text{V} \right] &\approx 0.85
\end{align}

These conditional probabilities are sometimes referred to as ``classical correlations'', because a classical two-particle state that has no entanglement could conceivably be constructed that would also yield such strong correlations. However, when these probabilities are considered in combination with the results shown in the right panel of Figure \ref{fig:nat_correlations}, they are unambiguously reflective of a two-qubit state that is entangled. The conditional probabilities in the orthogonal basis also show strong correlations; for example, $\text{Pr}\left[ \text{spin}=\leftarrow \middle| \text{photon}=\sigma^+ \right] \approx 1$. These measured conditional probabilities are strikingly similar to those we would expect if there were no measurement errors, and the state we produced was $\ket{\psi} = \frac{1}{\sqrt{2}} \left( \ket{\uparrow} \ket{i \text{H}, \hbar ( \omega + \delta_\text{e} ) } + \ket{\downarrow} \ket{\text{V}, \hbar \omega } \right)$. 

This is easy to see if we rewrite the state $\ket{\psi}$ using the $\left\{ \ket{\rightarrow} , \ket{\leftarrow} \right\}$ basis for the spin, and the $\left\{ \ket{\sigma^+} , \ket{\sigma^-} \right\}$ basis for the photon:

\begin{align}
  \ket{\psi} &= \frac{1}{\sqrt{2}} \left( \ket{\uparrow} \ket{i \text{H} } + \ket{\downarrow} \ket{\text{V}} \right) \\
	           &= \frac{1}{\sqrt{2}} \left( i \ket{\uparrow} \ket{\text{H}} + \ket{\downarrow} \ket{\text{V}} \right) \label{eq:SpinPhotonEntanglement_noenergy} \\
						 &= \frac{i}{\sqrt{2}} \left( \ket{\rightarrow} \ket{\sigma^-} + \ket{\leftarrow} \ket{\sigma^+} \right) \label{eq:SpinPhotonEntanglement_orthbasis}
\end{align}

Here we have neglected the energy information, since in De Greve {\it et al.} \cite{de_greve_quantum-dot_2012,de_greve_complete_2013} (and in the work by Schaibley {\it et al.} \cite{schaibley_demonstration_2013}), the energy information is not measured.\footnote{In Refs. \cite{de_greve_quantum-dot_2012,de_greve_complete_2013}, due to the sub-$10$-ps timing resolution achieved using pulsed downconversion, energy information that can distinguish between the two photons is unobtainable even in principle. Explicit ``erasure'' of the energy information is crucial; simply not measuring would lead to tracing over all possible outcomes, resulting in a reduction of the observed state fidelity.} A similar rewriting procedure is used for frequency/energy photonic qubits instead of polarization qubits in Gao {\it et al.} \cite{gao_observation_2012}. 

Equation \ref{eq:SpinPhotonEntanglement_orthbasis} indicates that we should expect the conditional probabilities in the orthogonal bases to be $\text{Pr}\left[ \text{spin}=\rightarrow \middle| \text{photon}=\sigma^- \right] = 1$, $\text{Pr}\left[ \text{spin}=\leftarrow \middle| \text{photon}=\sigma^- \right] = 0$, $\text{Pr}\left[ \text{spin}=\rightarrow \middle| \text{photon}=\sigma^+ \right] = 0$, and $\text{Pr}\left[ \text{spin}=\leftarrow \middle| \text{photon}=\sigma^+ \right] = 1$. 

An important subtlety in these experiments \cite{de_greve_quantum-dot_2012,de_greve_complete_2013,schaibley_demonstration_2013} arises from the fact that as soon as the photonic qubit is measured in the $\left\{ \ket{\sigma^+} , \ket{\sigma^-} \right\}$ basis, the spin state collapses to either $\ket{\rightarrow}$ or $\ket{\leftarrow}$ (depending on the photon polarization measurement outcome), and due to the presence of an external magnetic field, the spin will undergo Larmor precession. For example, if the spin state is collapsed to $\ket{\rightarrow}$, after half a Larmor period, it will have evolved to become $\ket{\leftarrow}$. This Larmor precession is a convenient feature, since when it is combined with optical rotation pulses, it allows for the measurement of the spin in bases other than $\left\{ \ket{\uparrow} , \ket{\downarrow} \right\}$. However, it also has a detrimental effect: these experiments are performed on time ensembles, where the same quantum dot is observed many times, and in each run of the experiment, the spontaneous decay of the $\ket{\uparrow\downarrow\Downarrow}$ state can occur at a different time (roughly within the lifetime of that trion state, which was approximately $600~\ps$ in Refs. \cite{de_greve_quantum-dot_2012,de_greve_complete_2013}). The timing resolution of the detection used to measure the photonic qubit is thus crucial; if the timing resolution is not much faster than the Larmor period, the experimeter will bin together runs of the experiment where the trion decays occurred at substantially different times\footnote{The ``time'' here means the time delay between a synchronization pulse that occurs at the start of every run of the experiment, and of the photon emission by the quantum dot, as opposed to the absolute time.}, and the spin measurements in the $\left\{ \ket{\rightarrow} , \ket{\leftarrow} \right\}$ basis will consequently yield greatly reduced correlations. This is explained in detail in the supplementary information of Reference \cite{de_greve_quantum-dot_2012}. The solution used in the experiments reported in References \cite{de_greve_quantum-dot_2012,de_greve_complete_2013} was to develop an ultrafast (sub-10-ps) optical gate, using frequency downconversion, which resulted in an effective timing resolution of photon detection of approximately $8~\ps$. This compared to a Larmor period of approximately $57~\ps$. This technique provided the added benefit that the frequency conversion that was performed had a target wavelength of approximately $1560~\nm$, which is in the low-loss band used for telecommunications in optical fibres. 

\begin{figure}[t]
\centering
\includegraphics[width=12cm]{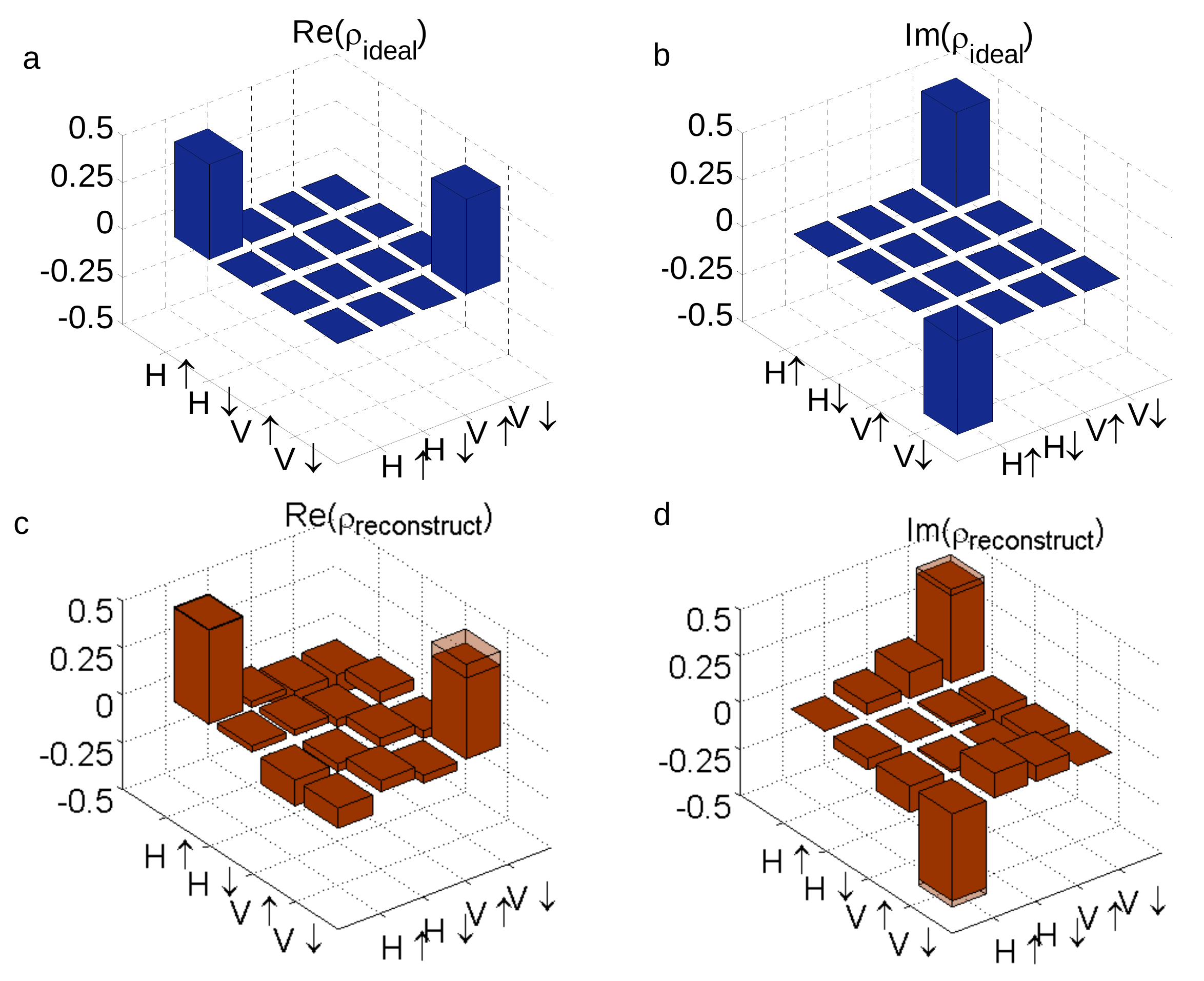}
\caption{Entangled Spin-Photon State Density Matrices. Reprinted with permission from De Greve {\it et al.} \cite{de_greve_complete_2013}. (a) The real and (b) imaginary parts of the ideal density matrix $\rho_\text{ideal}$. (c) The real and (d) imaginary parts of the density matrix reconstructed using the direct procedure, $\rho_\text{reconstruct}$. The shaded regions depict the ideal density matrix.}
\label{fig:natcomms_fig2}
\end{figure}

The ideal density matrix is $\rho_\text{ideal} = \ket{\psi}\bra{\psi}$, where $\ket{\psi}$ is given in Eq. \ref{eq:SpinPhotonEntanglement_noenergy}. A two-qubit ($4 \times 4$) density matrix has 16 complex entries, but since it is constrained to be Hermitian, it can be specified by just 16 real numbers.\footnote{Since a density matrix should also have a trace of one, a two-qubit density matrix should only need 15 real numbers to be specified, although typically in quantum state tomography 16 numbers are used, since the reconstruction procedure is more convenient in this case.} As is described in detail in the supplementary information of Ref. \cite{de_greve_complete_2013}, De Greve {\it et al.} performed measurements to obtain 16 conditional probabilities in a combination of three different orthogonal bases for the spin and the photon polarization. The reconstructed density matrix can be computed using the formula $\rho_\text{reconstruct} = \frac{1}{4} \sum_{i,j} r_{i,j} \sigma_i \otimes \sigma_j$, where the $r_{i,j}$ are related to the measurement results ($r_{i,j} \triangleq \text{Tr}\left[ \rho \sigma_i \otimes \sigma_j \right]$) \cite{nielsen_quantum_2011}. The ideal and reconstructed density matrices are depicted in Figure \ref{fig:natcomms_fig2}, in the $\left\{ \ket{\text{H}} , \ket{\text{V}} \right\} \otimes \left\{ \ket{\uparrow} , \ket{\downarrow} \right\}$ basis.

This direct reconstruction of the density matrix is simple, but has a flaw: due to imperfections in the measurements (for example, detector dark counts), the reconstructed density matrix may be non-physical: it may not have a trace of one, and moreover, it may not be positive semi-definite. The trace can be forced to be one by normalizing the reconstructed density matrix, but there is no simple method to force the matrix to be positive semi-definite after it has already been reconstructed using the direct method. 

One can see an example of the kind of measurement error that results in a non-physical result in Figure \ref{fig:nat_correlations}: note that the conditional probability $\text{Pr}\left[ \text{spin}=\uparrow \middle| \text{photon}=\text{H} \right]$ is measured as being slightly greater than 1. This isn't physically possible, since the probability of measuring the spin to be in state $\ket{\uparrow}$ is at most 1. 

One solution to this problem that is commonly used in quantum state tomography is to perform a reconstruction of the density matrix that is constrained to produce the positive-semi-definite, trace-one density matrix that is most consistent the measurement results. This can be done using a maximum likelihood estimation (MLE) procedure, as described by James {\it et al.} \cite{james_measurement_2001}, and in the supplementary information of De Greve {\it et al.} \cite{de_greve_complete_2013}. The MLE procedure produces a density matrix that we denote as $\rho_\text{MLE}$. 

\begin{figure}[t]
\centering
\includegraphics[width=12cm]{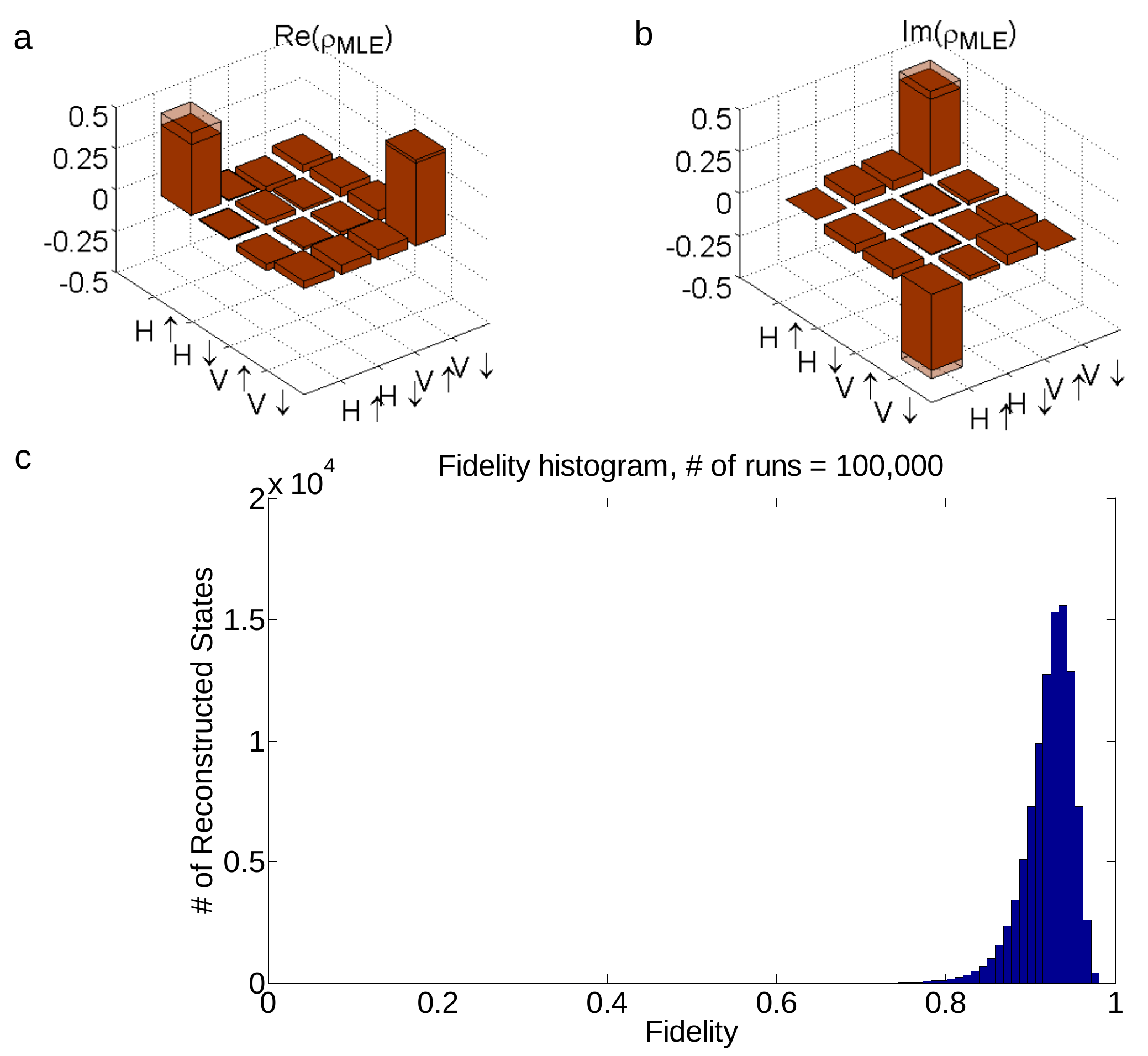}
\caption{Maximum-Likelihood-Estimation-based Density Matrix Reconstruction and Uncertainty Analysis. Reprinted with permission from De Greve {\it et al.} \cite{de_greve_complete_2013}. (a) The real and (b) imaginary parts of the density matrix reconstructed using the MLE procedure, $\rho_\text{MLE}$. The shaded regions depict the ideal density matrix. (c) Histogram of fidelities of reconstructed density matrices using resampled data. The mean and median fidelities are $92.1\%$ and $92.7\%$ respectively, and the standard deviation is $3.2\%$.}
\label{fig:natcomms_fig3}
\end{figure}

Since the procedure used to obtain $\rho_\text{MLE}$ is an iterative numerical optimization, it is not possible to use standard propagation of error methodology to determine the uncertainty in, for example, the fidelity $F$ of the state ($F \triangleq \left\langle \psi_\text{ideal} |\rho_\text{MLE} \middle| \psi_\text{ideal} \right\rangle$). However, by resampling \cite{efron_introduction_1993,de_greve_complete_2013} the original photon counting data, it is possible to generate a distribution of reconstructed density matrices, and hence a distribution of metrics on those matrices. Figure \ref{fig:natcomms_fig3} shows both the density matrix reconstructed using the MLE procedure on the original data, and the distribution of fidelities of the matrices obtained via resampling. 

The mean fidelity of the spin-photon entangled state produced by De Greve {\it et al.} in Ref. \cite{de_greve_complete_2013} was $F=92.1\%$, with a single-standard-deviation uncertainty of $\pm 3.2\%$. If two spatially-separated quantum dots are used to produce spin-photon entangled qubits each with fidelity greater than $1/\sqrt{2} \approx 0.71$, and we assume perfect photon interference, then a spin-spin entangled state can certainly be produced that will have a fidelity greater than $0.5$. Therefore a spin-photon state fidelity of $0.92$ is certainly sufficient for a demonstration of spin-spin entanglement with quantum dots.

Table \ref{tab:SpinPhotonComparison} compares this spin-photon state fidelity to that of fidelities obtained in other spin-photon entanglement experiments. This table also highlights the feature that quantum dots have optically excited states with relatively short lifetimes ($\sim 600~\ps$ when the quantum dot is embedded in low-$Q$ planar microcavity); this affects the rate at which spin-photon entangled states can be generated.

\begin{table}
\caption{Spin-Photon Entanglement Results. A summary of spin-photon entanglement generation experiment results in different single-particle physical systems, given in chronological order.}
\label{tab:SpinPhotonComparison}
\begin{tabular}{p{3.3cm}p{1.8cm}p{1.7cm}p{1.95cm}p{2cm}p{1.2cm}}
\hline\noalign{\smallskip}
Physical System & Photon Emission Wavelength & Spontaneous Emission Time & Entangled State Fidelity & Institution & Reference \\
\noalign{\smallskip}\hline\noalign{\smallskip}
Trapped Ion ($^{111}\text{Cd}^+$) & $214.5 \nm$ & $3 \ns$ & $\geq 87\%$ & U. Maryland$^a$ & \cite{blinov_observation_2004}\\
Neutral Atom ($^{87}\text{Rb}$) & $780 \nm$ & $26 \ns^b$ & $87 \pm 1 \%$ & LMU M\"unchen & \cite{volz_observation_2006}\\
Neutral Atom ($^{87}\text{Rb}$) & $780 \nm$ & $26 \ns^c$ & $> 86.0(4) \%$ & MPI Garching & \cite{wilk_single-atom_2007}\\
NV Centre & $637 \nm$ & $11 \ns$ & $\geq 70 \pm 7 \%$ & Harvard U. & \cite{togan_quantum_2010}\\
Trapped Ion ($^{40}\text{Ca}^+$) & $854 \nm$ & $14 \ns$ & $97.4 \pm 0.2 \%$ & U. Innsbruck & \cite{stute_tunable_2012}\\
Quantum Dot (InAs) & $910 \nm \newline \rightarrow 1560 \nm^d$ & $0.6 \ns$ & $\geq 80 \pm 8.5 \%$ & Stanford U. & \cite{de_greve_quantum-dot_2012}\\
Quantum Dot (InGaAs) & $967 \nm$ & $\approx 1 \ns$ & $\geq 68 \pm 5 \%$ & ETH Z\"urich & \cite{gao_observation_2012}\\
Quantum Dot (InAs) & $950 \nm$ & $\approx 1 \ns$ & $\geq 59 \pm 4\%$ & U. Michigan & \cite{schaibley_demonstration_2013}\\
Quantum Dot (InAs) & $910 \nm \newline \rightarrow 1560 \nm^d$ & $0.6 \ns$ & $92.1 \pm 3.2 \%$ & Stanford U. & \cite{de_greve_complete_2013}\\
\noalign{\smallskip}\hline\noalign{\smallskip}
\end{tabular}
\raggedright
$^a$ The group of Monroe was based at the University of Michigan at the time Ref. \cite{blinov_observation_2004} was published, but has since moved to the University of Maryland.\\
$^b$ Ref. \cite{hofmann_heralded_2012} (also from the group of Weinfurter at LMU) shows time-resolved measurements of $^{87}\text{Rb}$ $5^2\text{P}_{3/2} \rightarrow 5^2\text{S}_{1/2}$ decay that are consistent with the value of $26 \ns$ given in Ref. \cite{volz_precision_1996}.\\
$^c$ Ref. \cite{wilk_single-atom_2007} gives the atomic dipole decay rate as $2\pi \cdot 3 \MHz$ (HWHM). This is $2\pi \cdot 6 \MHz$ (FWHM), and corresponds to a $\approx 26 \ns$ lifetime, as per Ref. \cite{volz_precision_1996}. \\
$^d$ The quantum dot emitted photons at $910 \nm$, but these photons were converted to the telecommunications wavelength, $1560 \nm$.
\end{table}

\section{Conclusion}

We have explained how quantum dots might be used as the building blocks for a quantum repeater, but there is still much work to be done before a useful quantum repeater may be built from quantum dots, or indeed before such a repeater can even be designed in detail. 

In the short term, one of the major outstanding experimental goals is the demonstration of spin-spin entanglement using quantum dots, i.e., the entanglement of spins in two different quantum dots that are spatially separated by a macroscopic distance. Spin-spin entanglement has been achieved using atomic ensembles \cite{chou_measurement-induced_2005,chou_functional_2007}, trapped ion qubits \cite{moehring_entanglement_2007}, single atom qubits \cite{hofmann_heralded_2012,ritter_elementary_2012}, and NV center qubits \cite{bernien_heralded_2013}, so the spin-spin generation protocols are well-tested, but a demonstration of spin-spin entanglement with quantum dots is nevertheless seen as an important milestone for the quantum dot spin qubit community. 

The spin readout mechanism that was used in all the recent quantum dot spin-photon entanglement experiments we have highlighted \cite{de_greve_quantum-dot_2012,gao_observation_2012,schaibley_demonstration_2013,de_greve_complete_2013} yields only a single photon (at most) per experimental run, and so can only be used as a multi-shot readout by averaging over many experimental runs (since photon collection and detection efficiency is not unity). Demonstration of a single-shot readout mechanism that can be integrated with the other important operations for a quantum repeater memory qubit is an important goal. This would make a spin-spin demonstration easier (since then only two-photon coincidences would need to be observed, rather than four-photon coincidences), and is also a key requirement for implementation of the surface code \cite{jones_layered_2012}. 

The lack of a scalable two-qubit gate is arguably the biggest challenge that the community needs to overcome. Two-qubit operations on memory qubits are ubiquitous in all the large-scale quantum repeater proposals we have discussed, and at the very least will be needed to perform error correction. There have as-yet been no demonstrations in any physical system of quantum error correction that allow a memory qubit to stay coherent for an arbitrarily long time, or even for a time much longer than the native $T_2$ time. However, with quantum dot spin qubits, this is especially important even for early demonstrations, since InAs quantum dot spin qubits have rather short $T_2$ times, which limit the communication distance. Demonstrating a QEC-enabled extension of a logical qubit coherence time in quantum dots is a major goal, but one that can only be tackled after a scalable two-qubit gate has been developed. 

The high-level designs for large-scale quantum devices using quantum dots call for the use of arrays of site-controlled quantum dots, but here too there is much work to be done: developing methods to produce such arrays with a high yield of quantum dots that have good, homogeneous optical properties is a major challenge. In the near term, spin results that have been achieved using randomly-located quantum dots should be replicated using site-controlled quantum dots, to aid in the development of site-controlled QD arrays that are suitable for spin qubits. 

The challenges in constructing a high-fidelity, high-bandwidth (measured in ``entangled qubit pairs per second'') quantum network are daunting. For approaches using neutral atoms or ions to succeed, researchers need to overcome significant barriers to scaling. Meanwhile solid-state approaches have struggled to achieve the required operation fidelities for fault-tolerant operation, in some cases suffer from insufficiently-long coherence times, and in many cases don't yet have a scalable two-qubit gate operation, among other imperfections. Gisin and Thew, in 2007 \cite{gisin_quantum_2007}, wrote: 

\begin{quotation}
``The development of a fully operational quantum repeater and a realistic quantum-network architecture are grand challenges for
quantum communication. Despite some claims, nothing like this has been demonstrated so far and one should not expect any real-world
demonstration for another five to ten years.''
\end{quotation}

Seven years later, much the same can still be said. Several months after Gisin and Thew's review was published, results showing entanglement between distant quantum memories made from atomic ensembles \cite{chou_functional_2007}, and between single trapped ions \cite{moehring_entanglement_2007} were reported. As we have already mentioned, many other experiments generating entanglement between a quantum memory and a photon, or between quantum memories, have subsequently been performed \cite{togan_quantum_2010,stute_tunable_2012,hofmann_heralded_2012,ritter_elementary_2012,de_greve_quantum-dot_2012,gao_observation_2012,schaibley_demonstration_2013,de_greve_complete_2013,bernien_heralded_2013}. However, even the demonstration of just a single round of entanglement purification between distant quantum memories has not yet been completed, in any physical system. Similarly, there have been no demonstrations of entanglement swapping to connect two distant quantum memories via an intermediate quantum memory. Entanglement generation between two quantum memories is now well-established in both atomic and in some solid-state systems, but the goal of implementing both entanglement swapping and entanglement purification between remote quantum memories to demonstrate a proof-of-concept quantum repeater is still a distant hope rather than a soon-to-be-completed milestone.

Building a quantum repeater is evidently a grand challenge, and one that seems unlikely to be met for many years to come. Besides the challenges for how to make a quantum repeater, there is also a challenge to find uses for a quantum repeater. The canonical application at present is long-distance quantum key distribution. However, private key distribution can currently be performed with very high bandwidth using classical means, and the cost-benefit analysis for quantum repeaters for this application is not necessarily favourable.\footnote{A briefcase packed with hard drives or tapes can easily store over $100~\text{TB}$ of private keys, and can be transported to the opposite side of the world via a trusted courier in an airplane in approximately 24 hours. This yields a key distribution rate of approximately $10~\text{Gbits/sec}$. This can be compared to an achievable rate of Mbits/sec using $10^4$ quantum repeater stations, which was calculated by Fowler {\it et al.} \cite{fowler_surface_2010}. A hybrid approach has been suggested by Devitt {\it et al.} \cite{devitt_high-speed_2014}: if quantum memories can be made to have coherence times on the order of weeks, then Devitt {\it et al.} propose literally shipping the quantum memories from Alice to Bob. This avoids the need for long-distance fibre communication. The only use of optics in such an implementation of long-distance entanglement distribution would likely be the initial generation of the entanglement over a distance of a few meters. The coverage in this chapter of entanglement generation and construction of quantum memories, both in the abstract and for the particular case of quantum dots, is still relevant for such a physical-transport-based entanglement distribution scheme.} Tests of the Bell inequality over ever-longer distances are an interesting fundamental application of the distributed entanglement that quantum repeaters would provide. Teleportation \cite{bennett_teleporting_1993} is also an interesting fundamental application of distributed entanglement, and may also play a role in the construction of distributed quantum computers \cite{van_meter_architecture_2006}. Gottesman {\it et al.} \cite{gottesman_longer-baseline_2012} have proposed an optical interferometer design that could overcome current optical telescope resolution limits if a quantum repeater is realized, and K\'om\'ar {\it et al.} \cite{komar_quantum_2014} have proposed a global atomic clock network design that fundamentally uses remote entangled states. Both these proposals are recent, so there is some hope that more uses of distributed entanglement may yet be uncovered.

% =================================================================================================================

\section*{Acknowledgements} %
We would like to acknowledge our coauthors of those papers covered in this chapter that we were involved in (namely References \cite{press_ultrafast_2010,de_greve_quantum-dot_2012,de_greve_complete_2013}), with whom we had many valuable discussions. We would like to thank especially Yoshihisa Yamamoto (who supervised these works), Thaddeus Ladd, Christian Schneider and Cody Jones for numerous conversations about the place and use of quantum dots in quantum communication technology. We would also like to thank Ana Predojevi\'c for her thorough editing of this chapter, including many helpful suggestions, and Jelena Vu\v{c}kovi\'c for her careful reading of a draft of this chapter. PLM acknowledges support from the Cabinet Office, Government of Japan, and the Japan Society for the Promotion of Science (JSPS) through the Funding Program for World-Leading Innovative R\&D on Science and Technology (FIRST Program). KDG acknowledges support from a Harvard Quantum Optics Center (HQOC) Fellowship.

\printbibliography

\end{document}

%% file: CustomCommands.tex
\RequirePackage{amsmath}
\RequirePackage{amssymb}
\RequirePackage{bm}
\RequirePackage{graphicx}
\RequirePackage{color}
\RequirePackage[active]{srcltx}
\RequirePackage{xparse}

\NewDocumentCommand{\sectionc}{o m o}
  {\IfNoValueTF{#3}% No short title for the header
     {\IfNoValueTF{#1}% No title for the toc
        {\section{#2}}
        {\section[#1]{#2}}%
     }
     {\IfNoValueTF{#1}% No title for the toc
        {\section[#2]{#2\sectionmark{#3}}\sectionmark{#3}}
        {\section[#1]{#2\sectionmark{#3}}\sectionmark{#3}}%
    }%
  }
\long\def\comment#1{}
\newcommand{\mathcommand}[3][0]{\newcommand{#2}[#1]{\ensuremath{#3}}}
\mathcommand{\radtwo}{\sqrt{2}}
\mathcommand[1]{\oneover}{\frac{1}{#1}}

\newcommand{\be}{\begin{equation}}
\newcommand{\ee}{\end{equation}}

\mathcommand{\eup}{\uparrow}
\mathcommand{\edown}{\downarrow}
\mathcommand{\tup}{\uparrow\downarrow,\Uparrow}
\mathcommand{\tdown}{\uparrow\downarrow,\Downarrow}

\mathcommand[1]{\supop}{\hat{#1}} % to promote an operator to a superoperator
\mathcommand[1]{\hilbert}{\mathfrak{H}_{\text{#1}}} %symbol for Hilbert space
\mathcommand[2]{\avehamt}{\ts{\tilde{\mathcal{H}}}{#1}^{(#2)}} % Used for average Hamiltonian terms in paper

\mathcommand[1]{\smallket}{|#1\rangle}
\mathcommand[1]{\smallbra}{\langle #1|}
\mathcommand[1]{\bigket}{\bigl|#1\bigr\rangle}
\mathcommand[1]{\bigbra}{\bigl\langle #1\bigr|}
\mathcommand[1]{\biggket}{\biggl|#1\biggr\rangle}
\mathcommand[1]{\biggbra}{\biggl\langle #1\biggr|}
\mathcommand[1]{\ket}{\left| #1 \right\rangle}  %a Ket
\mathcommand[1]{\lket}{\bigl| #1 \bigr)}  %a Liouville Ket
\mathcommand[1]{\bra}{\left\langle #1 \right|}  %a Bra
\mathcommand[1]{\lbra}{\bigl( #1 \bigr|}  %a Bra
\newcommand{\braket}[2]{\left\langle #1 |#2\!\right\rangle} %an inner product
\mathcommand[1]{\bbra}{\biggl\langle #1 \biggr|}
\mathcommand[1]{\bket}{\biggl| #1 \biggr\rangle}
\mathcommand[1]{\power}{\times 10^{#1}}
\mathcommand{\tesla}{\text{~T}}
\mathcommand{\cc}{\text{~cm}^3}
\mathcommand{\kelvin}{\text{~K}}
\mathcommand{\percc}{\text{~cm}^{-3}}
\mathcommand{\mol}{\text{~mol}}

\mathcommand{\per}{/\!\!} %Gets rid of that hard space

\mathcommand{\rad}{\text{~rad}}

\mathcommand{\joule}{\text{~J}}
\mathcommand{\Joule}{\text{~J}}
\mathcommand{\meV}{\text{~meV}}
\mathcommand{\ueV}{\text{~}\mu\text{eV}}
\mathcommand{\eV}{\text{~eV}}
\mathcommand{\keV}{\text{~keV}}
\mathcommand{\MeV}{\text{~MeV}}
\mathcommand{\GeV}{\text{~GeV}}
\mathcommand{\uW}{\text{~}\mu\text{W}}
\mathcommand{\watts}{\text{~W}}
\mathcommand{\watt}{\watts}
\mathcommand{\volts}{\text{~V}}
\mathcommand{\mHz}{\text{~mHz}}
\mathcommand{\Hz}{\text{~Hz}}
\mathcommand{\kHz}{\text{~kHz}}
\mathcommand{\MHz}{\text{~MHz}}
\mathcommand{\GHz}{\text{~GHz}}
\mathcommand{\THz}{\text{~THz}}
\mathcommand{\PHz}{\text{~PHz}}
\mathcommand{\yr}{\text{~yr}}
\mathcommand{\hr}{\text{~hr}}
\mathcommand{\minutes}{\text{~min}}
\mathcommand{\seconds}{\text{~sec}}
\mathcommand{\second}{\text{~sec}}
\mathcommand{\ms}{\text{~ms}}
\mathcommand{\us}{\text{~}\mu\text{s}}
\mathcommand{\ns}{\text{~ns}}
\mathcommand{\ps}{\text{~ps}}
\mathcommand{\fs}{\text{~ps}}
\mathcommand{\km}{\text{~km}}
\mathcommand{\meter}{\text{~m}}
\mathcommand{\cm}{\text{~cm}}
\mathcommand{\mm}{\text{~mm}}
\mathcommand{\um}{\text{~}\mu\text{m}}
\mathcommand{\nm}{\text{~nm}}
\mathcommand{\picom}{\text{~pm}}
\mathcommand{\fm}{\text{~fm}}
\mathcommand{\angstr}{\text{~\AA}}
\mathcommand{\gram}{\text{~g}}
\mathcommand{\kg}{\text{~kg}}

\mathcommand{\mub}{\ts{\mu}{B}}  % Bohr Magneton
\mathcommand{\kb}{\ts{k}{B}}     % Boltzmann's constant
\mathcommand{\ef}{\ts{E}{F}}     % The Fermi Energy

\mathcommand{\dB}{\text{~dB}}